\documentstyle[12pt,axodraw]{article}
\def\mysection#1{\refstepcounter{section}\subsection{#1}
                 \setcounter{equation}{0}}
\def\mysubsection#1{\subsubsection{#1}}
\def\theequation{\thesection.\arabic{equation}}
\def\thefigure{\thefigure.\arabic{equation}}

\def\cD{{\cal D}}
\def\cR{{\cal R}}
\def\a{\alpha}
\def\b{\beta}
\def\d{\delta}
\def\De{\Delta}
\def\ga{\gamma}
\def\la{\lambda}
\def\ka{\kappa}
\def\m{\mu}
\def\n{\nu}

\def\s{\sigma}

\def\th{\theta}

\def\ee{\varepsilon}
\def\gm{\Gamma}
\def\om{\omega}
\def\bareta{\bar{\eta}}
\def\La{\Lambda}
\def\Om{\Omega}
\def\oa{\rm 1a}
\def\ob{\rm 1b}
\def\YM{\rm YM}
\def\GF{\rm GF}

\def\const{{\rm const}}
\def\MM{\scriptscriptstyle M}
\def\WW{\scriptscriptstyle R}
\def\VV{\scriptscriptstyle V}
\def\cv{c_{\VV}}
\def\idq{\int\! {d^4\!q \over (2\pi)^4} \,\,}
\def\iddq{{1\over \m^{D-4}}\int\! {d^D\!q \over (2\pi)^D}\,\,}
\def\iddqgcv{{g^2\cv\over \m^{D-4}}\int\! {d^D\!q \over (2\pi)^D}\,\,}
\def\idqreg{\int_{\cR} {d^4\!q \over (2\pi)^4} \,\,}
\def\idx{\int\! d^4\!x \,\,}
\def\idy{\int\! d^4\!y \,\,}
\def\sumj{\sum_{j=1}^J}
\def\sumi{\sum_{i=1}^I}
\def\poLa{{p^2\over \La^2}}
\def\poMj{{p^2\over M_j^2}}
\def\pomi{{p^2\over m_i^2}}
\def\kaoLa{{\ka^2\over \La^2}}
\def\kaoMj{{\ka^2\over M_j^2}}
\def\kaomi{{\ka^2\over m_i^2}}
\def\ds{\displaystyle}
\def\ss{\scriptstyle}
\def\ie{{\it i.e.}}
\def\eg{{\it e.g.}}
\def\to{\rightarrow}
\def\VEV#1{\left\langle #1 \right\rangle}
\def\gtwo{{g^2 \cv \over 16\pi^2}}
\def\gim{{g^2 \cv \over 16\pi}}
\def\barc{\bar{c}}
\font\twelveBB=msym10 at 12pt
\newfam\BBfam
\textfont\BBfam=\twelveBB
\def\BB{\fam\BBfam\twelveBB}

\begin{document}


\begin{titlepage}
\rightline{FTUAM 94/9}
\rightline{NIKHEF-H 94/24}
\vskip 2 true cm
\begin{center}
  {\Large {\bf Higher covariant derivative Pauli-Villars}}\\
  \vskip 0.3 true cm
  {\Large {\bf regularization does not lead to a consistent QCD }}\\
  \vskip 1.2 true cm
  {\rm C. P. Martin}\\
  \vskip 0.3 true cm
  {\it Departamento de F\'\i sica Te\'orica, C-XI,
       Universidad Aut\'onoma de Madrid,} \\
  {\it Cantoblanco, Madrid 28049, Spain}\\
  \vskip 1.2 true cm
  {\rm F. Ruiz Ruiz}\\
  \vskip 0.3 true cm
  {\it NIKHEF-H, Postbus 41882, 1009 DB Amsterdam, The Netherlands}\\
\end{center}
\vskip 1 true cm
{\leftskip=1.5 true cm \rightskip=1.5 true cm
\noindent
We compute the beta function at one loop for Yang-Mills theory using
as regulator the combination of higher covariant derivatives and
Pauli-Villars determinants proposed by Faddeev and Slavnov. This
regularization prescription has the appealing feature that it is
manifestly gauge invariant and essentially four-dimensional. It
happens however that the one-loop coefficient in the beta function
that it yields is not $-11/3,$ as it should be, but $-23/6.$ The
difference is due to unphysical logarithmic radiative corrections
generated by the Pauli-Villars determinants on which the
regularization method is based. This no-go result discards the
prescription as a viable gauge invariant regularization, thus solving
a long-standing open question in the literature. We also observe that
the prescription can be modified so as to not generate unphysical
logarithmic corrections, but at the expense of losing manifest gauge
invariance.
\par}
\end{titlepage}
\setcounter{page}{2}

\mysection{Introduction}

To regularize and renormalize a theory with symmetries, it is very
convenient to use a regularization method and a renormalization scheme
that preserve the symmetries of the theory. In the case of gauge
theories, there are not so many regularization methods that preserve
gauge invariance. The two traditional candidates are dimensional
regularization \cite{tHooft} and higher covariant derivatives
\cite{Slavnov} \cite{Faddeev} --more recent regularization
methods \cite{Bern} \cite{Asorey} will not concern us here.
Dimensional regularization works well for vector gauge theories like
Yang-Mills or QCD but presents some problems when it comes to dealing
with theories with $\ga^5.$ As regards higher covariant derivative
regularization, it has the appealing feature that it does not have
problems with $\ga^5,$ thus constituting an in principle
useful tool to study anomalies. Furthermore, there are suggestions
\cite{SUSY} that it leads to supersymmetric as well as gauge invariant
regularization procedures. Yet it is not known whether higher
covariant derivative regulators work even for the simplest cases, like
ordinary Yang-Mills or QCD. In this paper we provide an answer to this
question in the negative for the case of Yang-Mills theory.

So let us consider Yang-Mills theory. Higher covariant derivative
regularization as proposed in ref. \cite{Faddeev} combines a higher
covariant derivative term $(D^2F)^2/4\La^4$ with a set of gauge
invariant Pauli-Villars determinants. There is some controversy in the
literature as to whether or not this prescription works. On the one
side there is the position defended in ref. \cite{Faddeev}, that
claims that the prescription works to all orders in perturbation
theory without problems. On the other there is the position of ref.
\cite{Warr}, that states that the prescription only regularizes
one-loop divergences and leaves two and higher loops unregularized.
Even worse, ref. \cite{Seneor} sustains that already at one loop the
divergences that arise in the regularized Green functions when the
regulators are removed can not be subtracted in a gauge invariant
fashion, so that the resulting renormalized theory is not gauge
invariant. This conclusion is in turn somewhat in contradiction with
the non-multiplicative renormalization schemes discussed in ref.
\cite{Day}. The problem is that many of these claims, if not all, are
based on rather qualitative arguments and are not supported by
explicit calculations, thus making it difficult to conclude about the
viability of the prescription. To the best of our knowledge, nowhere
in the literature physical issues like unitarity or if the
prescription reproduces asymptotic freedom with the correct beta
function are addressed.

In this paper we use the prescription to explicitly regularize and
renormalize Yang-Mills theory at one loop and to compute the beta
function at this order in perturbation theory. It turns out that the
renormalized theory is gauge invariant, thus giving lie to ref.
\cite{Seneor}, but it also turns out that the Pauli-Villars
determinants generate unphysical logarithmic radiative corrections.
These unphysical corrections modify the beta function of the theory so
that the one-loop coefficient in the beta function comes out to be
$-23/6$ and not $-11/3,$ as it should. In Minkowski space these extra
corrections change the imaginary part of some 1PI Green functions.
This is in contradiction with general results from renormalization
theory according to which the one-loop imaginary part of a 1PI Green
function is independent of the renormalization procedure. All this
implies that the resulting renormalized theory is inconsistent and
that the prescription is not an acceptable one and must be
disregarded.

The paper is organized as follows. In section 2 we review the
regularization prescription. We spend some time on this since there
are a few issues which have passed unnoticed in the previous
literature and they play a crucial part in the outcome of this
investigation. In section 3 we show that it is possible to subtract
the divergences from the regularized theory and have at the same time
gauge invariance. In section 4 we compute the beta function at one
loop and obtain that it does not have the correct coefficient. Section
5 is devoted to prove that the source of the unwanted contributions to
the beta function is the Pauli-Villars determinants on which the
prescription is based. We include two appendices which contain the
Feynman rules (Appendix A) and calculational details (Appendix B).

\medskip

\mysection{The regularization method}

Let us start introducing our notation and conventions. The theory we
are interested in is Yang-Mills theory in four-dimensional Euclidean
space. The classical action in a general covariant gauge has the
form
\begin{equation}
  S = S_{\YM} + S_{\GF} \>,
\label{action}
\end{equation}
where $S_{\YM}$ and $S_{\GF}$ are given by
\begin{equation}
  S_{\YM} = {1\over4} \> \idx F_{\m\n}^a F^{a\m\n}
\label{YMaction}
\end{equation}
and
\begin{equation}
  S_{\GF} \,= \idx \bigg[\> {\a\over 2}\,b^a b^a
          - b^a\,(\partial_\m A^{a\m})
          + \barc^a (\partial_\m D^\m c)^a \, \bigg] \>.
\label{GFaction}
\end{equation}
Here $\,F^a_{\m\n} = \partial_\m A^a_\n - \partial_\n A^a_\m + g
f^{abc} A^b_\m A^c_\n\,$ denotes the field strength, $A^a_\m$ the
gauge field, $g$ the coupling constant and $f^{abc}$ the structure
constants of the gauge algebra. The parameter $\a$ is the gauge-fixing
parameter, $b^a$ is the auxiliary field, $\barc^a$ and $c^a$ are the
Faddeev-Popov ghosts and $D_\m^{ac} = \d^{ac} + g f^{abc} A^b_\m$ is
the covariant derivative.  The Landau gauge corresponds to $\a=0,$ and
the Feynman gauge to $\a=1.$ We will assume that the gauge group is a
compact, simple Lie group so that the structure constants $f^{abc}$
can be taken completely antisymmetric in their indices without loss of
generality. Furthermore, we will normalize $f^{abc}$ so that
$f^{acd}f^{bcd}\!=\!\cv\d^{ab},$ with $\cv$ the quadratic Casimir
operator in the adjoint representation [$\cv=N$ for a gauge group
$SU(N)$]. After gauge fixing, gauge invariance takes the form of BRS
invariance and so the symmetry to keep under control is the BRS
symmetry. The BRS transformations read in our notation
\begin{equation}
  sA^a_\m = (D_\m c)^a \qquad sb^a =0 \qquad s\barc^a=b^a \qquad
  sc^a = -\,{1\over 2}\>g\, f^{abc} c^b c^c
\label{brstrans}
\end{equation}
with the BRS operator $s$ satisfying as usual $s^2\!=\!0.$ Finally,
the generating functional is formally given by
\begin{displaymath}
  Z\,[\,J,\chi,\eta,\bareta\,] \,=  \int \cD A ~ \cD b ~ \cD\barc
  ~ \cD c  ~ e^{-\,\left(\, S + S_J\,\right)} \>,
\end{displaymath}
where $S$ is as in eqs. (\ref{action})-(\ref{GFaction}), $S_J$ is
the source term
\begin{displaymath}
S_J \,= \idx \left( J^{a\m}\!A^a_\m + \chi^a b^a + \barc^a \eta^a
  + \bareta^a c^a \right)
\end{displaymath}
and $J^{a\m},~\chi^a,~\eta^a$ and $\bareta^a$ are sources for the
fields $A^a_\m,~b^a,~\barc^a$ and $c^a,$ respectively.

Since it will be used as a point of reference below, let us recall how
power counting goes. Given a 1PI Feynman diagram with $E_A$ external
gauge lines, $E_g$ external ghost lines and $L$ loops, its overall UV
degree $\om$ is given by
\begin{equation}
  \om = 4 - E_A - {3\over 2}\,E_g \>.
\label{powerYM}
\end{equation}
Superficially divergent 1PI diagrams have $\om\geq 0$ and correspond
to Green functions already present in the classical action $S.$ As is
well known, this ensures renormalizability. To actually renormalize
the theory, one proceeds in two steps. The first one is to provide a
regularization method that renders all Green functions finite order by
order in perturbation theory. The second one is to provide a
renormalization scheme that eliminates the divergences appearing in
the Green functions when the regulator is removed. In this paper we
use higher covariant derivative Pauli-Villars regularization as
proposed by Faddeev and Slavnov \cite{Slavnov} \cite{Faddeev} to
explicitly renormalize Yang-Mills theory at one loop. As announced in
the introduction, it will turn out that the resulting renormalized
theory is inconsistent. In the course of our investigation we have
found that there are a few issues concerning the regularization method
itself that have gone unnoticed in the previous literature and that
play a crucial role in the result. We thus find worth spending some
time in reviewing the method and bringing these points to light.

\mysubsection{Higher covariant derivative regulators}

The method proposed by Faddeev and Slavnov regularizes Yang-Mills
theory in two steps. In the first one, it introduces a higher
covariant derivative mass $\La$ by replacing the classical gauge-fixed
action $S$ in eq. (\ref{action}) with the action
\cite{Faddeev}
\begin{equation}
  S_\La \,=\, S_{\YM} + S_2 + S^f_{\GF} \>,
\label{actionLa}
\end{equation}
where $S_{\YM}$ is as in eq. (\ref{YMaction}) and $S_2$ and
$S^f_{\GF}$ are respectively given by
\begin{equation}
   S_2 = \, {1\over 4\La^4} \idx (D^2 F_{\m\n})^a (D^2 F^{\m\n})^a
\label{HCD2}
\end{equation}
and
\begin{displaymath}
  S^f_{\GF} = \idx \bigg[ \, {\a\over 2} ~ b^a
     {1\over f^2(\partial^2\!/\La^2)} \> b^a
          - b^a\,(\partial_\m A^{a\m})
          + \barc^a (\partial_\m D^\m c)^a \, \bigg] \>.
\end{displaymath}
Here $f(\partial^2\!/\La^2)$ is a polynomial in its argument
that in momemtum space becomes a function of $p^2/\La^2$ and that by
definition is chosen to satisfy the conditions
\begin{equation}
\begin{array}{c} {\ds
   f\!\left(p^2/\La^2 \right) \to 1
        \quad {\rm as} \quad \La\to\infty }\\[12pt] {\ds
   f\!\left(p^2/\La^2\right)\sim \left( p^2/\La^2 \right)^{\ga}
        \quad {\rm as} \quad |p|\to\infty \,, }
\end{array}
\label{fcondition}
\end{equation}
with $\ga\geq 1.$ In this paper we
will take
\begin{equation}
   f\!\left( \poLa \right) = \, 1 +\, {p^4 \over \La^4}
\label{ffunction}
\end{equation}
for reasons that we explain below. The propagator of the gauge field
for the action $S_\La$ has the form
\begin{equation}
   D^{ab}_{\m\n}(p) \,=\, \d^{ab}\,\bigg[\,
     {\La^4 \over p^4 \,(p^4+\La^4)} \>
     (\,p^2 g_{\m\n} - p_\m p_\n\,)\,
     + \,\a\> {p_\m p_\n \over p^4 f^2(p^2/\La^2)}\,\bigg] ~,
\label{propagatorLa}
\end{equation}
where $p^{2n}$ is a shorthand notation for $(p^2)^n.$ The second
condition in eq. (\ref{fcondition}) ensures that the propagator goes
as $1/p^6$ when $p^\m\to\infty.$ This and some power counting implies
that the overall UV degree $\om_\La$ of a 1PI diagram generated by
$S_\La$ with $E_A$ external gauge legs, $E_g$ external ghost legs and
$L$ loops is given by [compare with eq. (\ref{powerYM})]
\begin{equation}
   \om_\La = 4 - 4\,(L-1) - E_A - {7\over 2}\, E_g \>.
\label{powerLa}
\end{equation}
Now only one-loop 1PI diagrams with $E_A=2,3,4$ and $E_g=0$ are
superficially divergent. We thus see that the higher covariant
derivative regulator $\La$ improves the power counting behaviour of
the theory but does not completely regularize it. To achieve a
complete regularization, one must regularize the one-loop diagrams for
which $\om_\La\geq 0.$ Before doing so, let us make a few comments
that we feel are in order.

{\it Comment 1.} There are many possibles choices for $f$ compatible
with conditions (\ref{fcondition}). The choice in eq.
(\ref{ffunction}) ensures that $\a\!$-dependent contributions in any
1PI diagram are finite by power counting, as can be trivially checked.
In fact, any $f$ for which $\ga$ in eq. (\ref{fcondition}) is larger
or equal than 2 leads to superficially convergent $\a\!$-dependent
contributions. We have taken $f$ as in eq. (\ref{ffunction}) since in
this way the denominator of the propagator $D^{ab}_{\m\n}(p)$ has
factors only of type $p^4$ and $p^4+\La^4.$ Note also that we have
required $f$ to be polynomial so as to ensure locality.

{\it Comment 2.} Since the action $S_\La$ is BRS invariant, the higher
covariant derivative regulator $\La$ preserves BRS invariance.

{\it Comment 3.} Note that all 1PI diagrams generated by $S_\La$ with
external ghost legs are finite by power counting. This implies that
Yang-Mills 1PI diagrams with external ghosts are fully regularized by
the higher covariant derivative regulator $\La.$

{\it Comment 4.} One may wonder whether the theory can be completely
regularized by means of a higher covariant derivative term of higher
order in the covariant derivative. The answer is no. To see this,
consider the term
\begin{displaymath}
   S_n =\, {{\ss (-)}^{\,n}\over 4\La^{2n}} \>
          \idx F^a_{\m\n}\>(D^{2n}F^{\m\n})^a  \>,
\end{displaymath}
where $n$ is an arbitrary positive integer. Power counting now implies
that
\begin{displaymath}
   \om_\La(n) = 4 - 2n\,(L-1) - E_A
              - \left( {3\over 2} + n \right)g \>.
\end{displaymath}
So no matter how large $n$ is, $\om_\La(n)$ is always negative for
$L=1,$ $E_g=0$ and $E_A=2,3,4.$

\mysubsection{Pauli-Villars determinants}

To regularize the one-loop divergences in the Feynman diagram
expansion generated by the action $S_\La$ above, one can proceed as
follows \cite{Slavnov} \cite{Faddeev}. Consider the two following
generating functionals labeled by an index $r=0,1\!:$
\begin{equation}
  Z_r[\,J,\chi,\eta,\bareta\,] \,=
     \int \cD A ~\cD b ~ \cD \barc ~ \cD c ~
     e^{\,-\,\left(\, S_\La \,+\, S_J\,\right)} ~
     \prod_{j=1}^J \> \left( \det {\BB A}^r_j \right)^ {-\a_j/2} ~
     \prod_{i=1}^I \>  \left( \det {\BB C}^r_i \right)^{\ga_i}  \>,
\label{functional}
\end{equation}
Here $\{\a_j\}_J$ and $\{\ga_i\}_I$ are real parameters to be fixed
and $\,\det{\BB A}^r_j\,$ and $\,\det{\BB C}^r_i\,$ are defined by
\begin{equation}
   \left( \det {\BB A}^r_j\,\right)^{-1/2} =  \int \cD A_j\,\cD b_j
      ~ \exp\Big\{\!-\Big( S_{M_j} + S_{b_j} \Big)\Big\}
\label{detA}
\end{equation}
and
\begin{equation}
\det {\BB C}_i^r = \int \cD \barc_j \, \cD c_j ~
     \exp\Big\{\!-S_{m_i}\Big\} \>,
\label{detC}
\end{equation}
with
\begin{displaymath}
\begin{array}{c} {\ds
  S_{M_j} =\, {1\over2} \idx \! \idy  A_{j\m}^{~a}(x) \,\bigg[\,
     {\d^2 S_\La\over\d A^a_\m(x)\,\d A^b_\n(y)}\,-\,
     M_j^2\,\d^{ab} g^{\m\n} \d(x-y)\,\bigg]\, A_{j\n}^{~b}(y) }
\\[15pt] {\ds
  S_{b_j} = - \idx b^a_j \Big( \d^{ab} \partial_\m
            + rgf^{acb} A^c_\m \Big) \,A_j^{b\m} }
\\[12pt] {\ds
S_{m_i} = \idx \left[\,
        \barc_i^a\, \big( \partial_\m D^\m c_i\big){}^a
        + r\,g f^{abc}\barc_i^a\, A^b_\m \,(D^\m c_i){}^c
        -\, m_i^2 \, \barc_i^a c_i^a \,\right] \>.}
\end{array}
\end{displaymath}
The fields $\{A^{~a}_{j\m}\}$ are commuting Pauli-Villars fields of
mass $\{M_j\},$ the fields $b_j^a$ are Pauli-Villars Lagrange
multipliers imposing the condition
\begin{displaymath}
 \Big( \d^{ab} \, \partial_\m
      + r\,g\, f^{acb} A^c_\m \Big) A_j^{b\,\m} = 0
\end{displaymath}
and the fields $\{\barc^a_i,c^a_i\}$ are anticommuting Pauli-Villars
fields of mass $\{m_i\}.$ The determinants $\,\det{\BB A}_j^r$ and
$\,\det{\BB C}_i^r$ are BRS invariant if $r=1$ --actually gauge
invariant since they do not depend on ghosts. This follows from
the observation \cite{Faddeev} that the change of integration
variables $\phi \to \phi + \th (s\phi),$ where $\th$ is an
anticommuting parameter and $s\phi$ is given by
\begin{displaymath}
  sA^{~a}_{j\m} = g \, f^{abc}\, A^{~b}_{j\m}\, c^c \qquad
  sb_j^a = g\, f^{abc} \,b_j^b\, c^c \qquad
  s\barc^a_i = g \, f^{abc}\, \barc^b_i\, c^c \qquad
  sc^a_i = g \,f^{abc}\, c^b_i\, c^c \,,
\end{displaymath}
cancels the extra terms that arise in $\,\det{\BB A}_j^1$ and
$\,\det{\BB C}^1_i$ when the gauge field $A^a_\m$ is BRS-transformed.
However, if $r=0,$ the determinants $\,\det{\BB A}_j^r$
and $\,\det{\BB C}^r_i$ are not BRS invariant. Faddeev's and Slavnov's
proposal \cite{Slavnov} \cite{Faddeev} corresponds to taking $r=1$ and
is the one we will be studying in this paper. For later convenience,
though, we will keep $r$ explicit in our calculations.

The Feynman rules associated to the functional
$Z_r[J,\chi,\eta,\bareta]$ are listed in Appendix A. Note that the
determinant $\det{\BB A}^r_j$ has been defined in such a way that the
vertices $A^{2l+1}~(l=3,\ldots,6)$ have the same Feynman rule as
$A_j^2A^l.$ Note also that the Feynman rules for the vertices
$b_jAA_j,$ $\barc_iAc_i$ and $\barc_i A^2 c_i$ depend explicitly on
$r.$ Keeping these two observations in mind, we move on to discussing
how the Pauli-Villars determinants just introduced regularize the
one-loop divergences generated by $S_\La.$ We start discussing first
the case $r=0.$

\bigskip

\noindent {\underline {\sl Gauge non-invariant case: $r=0$}}

\medskip
\noindent
We want to show that the vacuum polarization tensor and the vertices
$\VEV{A^3}_{\rm 1PI}$ and $\VEV{A^4}_{\rm 1PI}$ at one loop generated
by $Z_0[J,\chi,\eta,\bareta]$ in eq. (\ref{functional}) are finite.
Let us start with the vacuum polarization tensor. At one loop it
receives contributions from the diagrams in Fig. 1. Let us look at the
sum of the diagrams in Figs. (1a) and (1b),
\begin{equation}
  \s_1 = {\rm (1a) + (1b)}\,.
\label{sum1}
\end{equation}
Recalling that $\a\!$-dependent contributions are superficially
convergent, we have that in diagram (1a) only the $\a=0$ part contains
divergences. Taking then $\a=0,$ noting that by construction the
vertices $A^3$ and $A^2A^2_j$ have the same Feynman rule, and using
the identity
\begin{equation}
  {1\over q^6+q^2\La^4+M_j^2\La^4} \,=\, {1\over q^2\,(q^4+\La^4)} \,
-\,{ M_j^2\La^4 \over q^2\,(q^4+\La^4)\>(q^6+q^2\La^4+M_j^2\La^4)}
\label{idsix}
\end{equation}
for the propagator of the Pauli-Villars field $A_j,$ it is trivial to
see that the only superficially divergent integrals in $\s_1$ are
quadratically divergent, do not depend on $M_j$ and pick an overall
factor $ 1+\sum_j \a_j.$ It thus follows that if we choose parameters
$\a_j$ such that
\begin{equation}
   1+\sumj \a_j = 0
\label{PValpha}
\end{equation}
the sum $\s_1$ becomes finite. In other words, diagram (1b)
regularizes diagram (1a) if condition (\ref{PValpha}) is satisfied. A
similar argument shows that the sum
\begin{equation}
   \s_2 = {\rm (1c)+(1d)}
\label{sum2}
\end{equation}
is finite if again condition (\ref{PValpha}) is met. Let us finally
consider the sum
\begin{equation}
 \s_3^0 = {\rm (1e) + (1f)} \,.
\label{sigma30}
\end{equation}
Noting that for $r=0$ the vertex $\barc_iAc_i$ has the same Feynman
rule as the vertex $\barc Ac$ and using the identity
\begin{displaymath}
  {1\over q^2 + m_i^2} \,=\, {1\over q^2} \,
      -\, {m_i^2 \over q^2 (q^2 + m_i^2)}
\end{displaymath}
twice, we find that $\s_3^0$ contains quadratic and logarithmic
divergences that do not depend on $m_i$ and that pick overall factors
$\,1+\sum_i\ga_i\,$ and $\,\sum_i\ga_im_i^2,$ respectively. Hence
imposing conditions
\begin{equation}
  1 + \sumi \ga_i = 0
\label{PVgamma}
\end{equation}
and
\begin{equation}
  \sumi \> \ga_i\,m_i^2 = 0\,,
\label{PVmass}
\end{equation}
we render the sum ${\rm (e)+(f)}$ finite.

To summarize, if the parameters $\a_j$ and $\ga_i$ and the masses
$M_j$ and $m_i$ satisfy the Pauli-Villars conditions
(\ref{PValpha})-(\ref{PVmass}), diagrams (1b), (1d) and (1f)
regularize diagrams (1a), (1c) and (1e) and the vacuum polarization
tensor becomes finite. Analogous arguments show that one-loop
divergences in the vertices $\VEV{A^3}_{\rm 1PI}$ and $\VEV{A^4}_{\rm
1PI}$ generated by $S_\La$ are regularized if the very same conditions
(\ref{PValpha})-(\ref{PVmass}) hold. This regularization procedure is
very neat, but as already pointed out is not gauge invariant.

\bigskip

\noindent {\underline {\sl Gauge invariant case: $r=1$}}

\medskip
\noindent
We know that the Feynman rules are not the same for $r=1$ as for
$r=0.$ Hence there is no reason to assume that the regularization
mechanism is the same in both cases. To study regularization for $r=1$
in detail, let us consider again the vacuum polarization tensor. At
one loop, it receives contributions from the diagrams in Fig. 1 and,
very importantly, from those in Fig. 2. We first look at the diagrams
in Fig. 1. Diagrams (1a) to (1d) are independent of $r,$ so the same
arguments as for $r=0$ imply that the sums $\s_1$ and $\s_2$ are
finite provided condition (\ref{PValpha}) holds. As concerns the sum
$\s_3^0,$ we note that diagram (1e) is independent of $r$ but (1f) is
not. Furthermore, the difference in between taking $r=1$ and $r=0$ in
diagram (1f) is a superficially divergent integral. Therefore, after
imposing conditions (\ref{PVgamma}) and (\ref{PVmass}), we are left
with unregularized divergences in $\s_3^0.$ Next we look at the
diagrams in Fig. 2, which we emphasize do not exist for $r=0.$ Some
trivial power counting shows that diagram (2a) is finite, while
diagrams (2b) to (2e) are superficially divergent. Hence, for the
generating functional $Z_1[J,\chi,\eta,\bareta]$ to produce a
finite vacuum polarization tensor, the divergences in (1e) and (1f)
must cancel the divergences in diagrams (2b) to (2e). In other words,
the sum
\begin{equation}
  \s_3^1 = {\rm (1e) + (1f) + (2b) + (2c) +(2d) + (2e)}
\label{sum31}
\end{equation}
must be finite. All this makes obvious that regularization for $r=1$
does not go as for $r=0.$ It also corrects some statements in the
previous literature \cite{Faddeev}, where it is explicitly claimed
that the terms that ensure manifest BRS invariance do not introduce
new divergences into the game.

Let us find the conditions on $\a_j,~\ga_i,~M_j$ and $m_i$ for which
the sum $\s_3^1$ is finite. A look at the topology of the diagrams
contributing to $\s_3^1$ reveals that an analysis of the type
performed for $\s_1$ and $\s_2$ is not possible. Yet we have to seek
for cancellations of divergences from different diagrams. To do this
with rigour, we introduce a regulator $\cR,$ compute the diagrams at
finite $\cR$ and see for what values of $\a_j,~\ga_i,~M_j$ and $m_i$
contributions divergent at $\cR=0$ cancel upon summation. To isolate
the divergences at $\cR=0$ in each one of the diagrams, we proceed as
follows. Consider for example diagams (2c) and (2d). With the choice
of momenta in Fig. 2 and after some Lorentz algebra\footnote{To
perform the Lorentz algebra of the diagrams in this paper, we have
used the algebraic package REDUCE \cite{reduce}}, we have
\begin{displaymath}
   {\rm (2c)+(2d)} = r g^2\cv \sumj \a_j\> \idqreg
       {N_{\m\n} \over (q+p)^2\, q^2\,(q^6+q^2\La^4+M_j^2\La^4)} ~,
\end{displaymath}
where
\begin{equation}
   N_{\m\n} = 2\, q^6\,(q_\m q_\n - q^2 g_{\m\n})
            + O\,(q^n,\,2\!\leq\!n\!<\!6)\>.
\label{Nmn}
\end{equation}
Terms in $N_{\m\n}$ of order five or less in $q^\m$ give rise to
integrals finite by power counting at $\cR=0.$ On the other hand,
terms in $N_{\m\n}$ of order six or higher in $q^\m$ give rise to
superficially divergent integrals at $\cR=0.$ This means that as far
as regularization is concerned it is enough to keep the terms written
explicitly in eq. (\ref{Nmn}). Using then the identities (\ref{idsix})
and
\begin{equation}
  {1\over q^4 + \La^4} \,=\, {1\over q^4} \,
                         -\, {1\over q^4 (q^4+\La^4)}
\label{idfour}
\end{equation}
and retaining only integrals which are superficially divergent at
$\cR=0,$ it is trivial to see that
\begin{displaymath}
  {{\rm (2c) + (2d)}\bigg\vert}_{\rm div} = - r\, g^2\cv\,\d^{ab} \,
     \sumj \, \a_j \> \idqreg { 2\, \big( g_{\m\n}- q_\m q_\n \big)
                           \over (p+q)^2} ~ .
\end{displaymath}
Proceeding in the same way for the other diagrams in $\s_3^1,$ summing
over diagrams and imposing condition (\ref{PValpha}), we obtain
\begin{equation}
\begin{array}{l} {\ds
  \s_3^1 (\cR)\,\bigg\vert_{\rm div} =  -\,g^2\cv\,\d^{ab} \,
     \bigg( 1 + \sumi \ga_i \bigg)
     \, \idqreg { 4\,q_\m q_\n + q_\m p_\n + p_\m q_\n
                  \over q^2 \> (q+p)^2 }   }\\[16pt]
\phantom{ {\ds \s^1_3 \Big\vert{\rm div} = -~~} } {\ds
     +\,2\,g^2\cv\,\d^{ab} \idqreg \bigg\{\,
     \sumi \ga_i \bigg[ \, {g_{\m\n} \over q^2 + m_i^2}
     + \, {4\,m_i^2\,q_\m q_\n \over q^4\,(q^2+m_i^2)} \,\bigg]\,
     + \, {g_{\m\n}\over (q+p)^2} \,
     \bigg\} }\\[16pt]
\phantom{ {\ds \s^1_3 \Big\vert{\rm div} = -~~}}  {\ds
     -\,g^2\cv\,\d^{ab} \, \sumi \, \ga_i \idqreg
     {q_\m p_\n + p_\m q_\n + p_\m p_\n \over q^2\>(q+p)^2}~. }
\end{array}
\label{sigma4}
\end{equation}
Assume we take as regulator $\cR$ dimensional regularization, \ie\
assume that
\begin{displaymath}
  \idqreg \,=\, \iddq   \qquad\quad D=4+2\ee \>,
\end{displaymath}
with $\m$ the dimensional regularization mass scale. Then, it is not
difficult to see that
\begin{equation}
  \s_3^1(\ee)\bigg\vert_{\rm div}
      = \gtwo\>\d^{ab} \,\bigg[ \, \bigg( 1+ \sumi \ga_i \bigg) \,
        \s_{\ee}\, \big( p^2 g_{\m\n} - p_\m p_\n\bigg)
      + \sumi \ga_i \,p^2 g_{\m\n} \,\bigg]
\label{pole}
\end{equation}
as $D\to 4,$ where
\begin{displaymath}
 \s_{\ee} =  {1\over 3}\, \bigg[ \, {1\over \ee}
     + \ln\!\bigg({p^2\over\m^2}\bigg) + \ga - {8\over 3}\,\bigg] ~.
\end{displaymath}
To get rid of the pole in eq. (\ref{pole}) and end up with a finte
$\s^1_3,$ we choose the parameters $\ga_i$ to satisfy the
Pauli-Villars condition (\ref{PVgamma}).

Putting everything together, we have that the regularized vacuum
polarization tensor is finite for parameters $\a_j$ and $\ga_i$
satisfying the Pauli-Villars conditions (\ref{PValpha}) and
(\ref{PVgamma}). Using the BRS identities and that the vacuum
polarization tensor, the ghost self-energy and the $\barc Ac\!$-vertex
are finite, it follows the three and four-point Green functions
$\VEV{A^3}_{\rm 1PI}$ and $\VEV{A^4}_{\rm 1PI}$ are also finite,
provided of course conditions (\ref{PValpha}) and (\ref{PVgamma})
hold. This closes the proof that the functional
$Z_1[J,\chi,\eta,\bareta]$ generates finite one-loop Green functions.
There are two important comments to be made at this point:

{\it Comment 1.} It is clear that regularization does not require
condition (\ref{PVmass}), contrarily to what is claimed in ref.
\cite{Faddeev} and all papers thereafter.

{\it Comment 2.} Divergences from isolated Feynman diagrams in
$\s_3^1$ get regularized only after integration at finite $\cR.$ This
is in contrast with what happens for $\s_1$ and $\s_2,$ where
regularization takes place algebraically prior to integration. Here we
have taken dimensional regularization as regulator $\cR$ because it
preserves BRS invariance. Assume that instead we take a cut-off $Q$
such that $|p|,m_i<<Q.$ Some elementary integration in eq.
(\ref{sigma4}) then leads to
\begin{equation}
  \s_3^1(Q)\Big\vert_{\rm div}
      = \gtwo\>\d^{ab} \,\bigg[ \bigg( 1+ \sumi \ga_i \bigg) \,
        \big( Q^2 g_{\m\n} + \s'_Q p^2 g_{\m\n}
                                     - \s_Q''p_\m p_\n\bigg)
      + \sumi \ga_i \bigg( p^2 g_{\m\n}
                      - {1\over 2} p_\m p_\n\bigg) \bigg]
\label{Qdiv}
\end{equation}
where
\begin{displaymath}
 \s'_Q = {1\over 3}\, \ln\!\bigg({Q^2\over p^2}\bigg) + {11\over 18}
 \qquad\quad
 \s''_Q = {1\over 3}\, \ln\!\bigg({Q^2\over p^2}\bigg) - {1\over 18}
 ~.
\end{displaymath}
Again we see that for $\ga_i$ satisfying eq. (\ref{PVgamma}), the sum
$\s_3^1$ is finite. Comparing eq. (\ref{pole}) with eq. (\ref{Qdiv}),
we note that the finite part that is left in $\s_3^1\big\vert_{\rm
div}$ after imposing condition (\ref{PVgamma}) is however different.
This implies that different $\cR{\rm's}$ give different local
contributions to $\s_3^1.$ As regards the non-local part of $\s_3^1,$
it is the same for all $\cR$ since it arises only from integrals which
are finite by power counting at $\cR=0.$

To sum up, Pauli-Villars determinants with $r=1$ are BRS invariant,
but checking that they provide the necessary regularization requires
introducing a pre-regulator $\cR.$ This means strictly speaking that
$r=1$ Pauli-Villars determinants do not provide an acceptable
regularization. Still one could argue that since non-local
contributions are independent of the choice of pre-regulator $\cR,$
$r=1$ Pauli-Villars regularization is good enough. In sections 4 and 5
we show that this is not either the case, since the determinants
$\,\det{\BB A}^1_j$ and $\,\det{\BB C}^1_i$ introduce unphysical
logarithmic radiative corrections.

\medskip

\mysection{The renormalized theory}

So far we have a BRS invariant regularization prescription that
involves the masses $\La,~M_j$ and $m_i$ (and a pre-regulator $\cR$
which from now on we take it to be dimensional regularization). The
next step is to define the renormalized theory. There are several
requirements that the resulting renormalized theory must satisfy for
it to make sense. One of them is that the renormalized Green functions
satisfy the BRS identities. To date, there is no agreement in the
literature \cite{Seneor} as whether starting from the generating
functional $Z_1[J,\chi,\eta,\bareta]$ it is possible to define
renormalized Green functions compatible with BRS invariance. In this
section we show that this is indeed possible at one loop in the least.

To study the BRS identities, it is convenient to work with the
effective action. The effective action associated to the functional
$Z_1[J,\chi,\eta,\bareta]$ is defined in two steps. One first
introduces external fields $K^{a\m}$ and $H^a$ coupled respectively to
the non-linear BRS transforms $sA^a_\m$ and $sc^a$ so that the
generating functional becomes
\begin{displaymath}
  Z_1[J,\chi,\eta,\bareta;K,H] \,=
     \int \cD A ~\cD b ~ \cD \barc ~ \cD c ~
     e^{\,-\,\left(\, S\La \,+\, S_J\,+\,S_{\rm ext}\,\right)} ~
     \prod_{j=1}^J \> \left( \det {\BB A}^1_j \right)^ {-\a_j/2} ~
     \prod_{i=1}^I \>  \left( \det {\BB C}^1_i \right)^{\ga_i}  \>,
\end{displaymath}
with $S_{\rm ext}$ given by
\begin{equation}
   S_{\rm ext} = \idx \Big[\, K^{a\m}(sA^a_\m)
          + H^a (sc^a) \Big] \>.
\label{extaction}
\end{equation}
And secondly, one writes
\begin{displaymath}
Z_1[J,\chi,\eta,\bareta;K,H\,]\,=\, \exp \Big( \!-
       W[J,\chi,\eta,\bareta;K,H]\,\Big)
\end{displaymath}
and performs a Legendre transformation on $W[J,\chi,\eta,\bareta;K,H]$
and on the sources $J^{a\m},$ $\!\chi^a,$ $\!\eta^a$ and $\bareta^a.$
The Legendre transform of $W[J,\chi,\eta,\bareta;K,H]$ is the
regularized effective action, a functional of mass dimension four and
ghost number zero that depends on the Legendre fields
$A^a_\m,b^a,\barc^a,c^a$ and on the external fields $K^{a\m},H^a$ and
that generates 1PI Green functions. In what follows we will denote by
the letter $\psi$ the set of fields
$\{\psi\}=\{A^a_\m,b^a,\barc^a,c^a, K^{a\m},H^a\}$ and write $\gm_{\La
M_j m_i}[\psi]$ for the regularized effective action, our conventions
for the mass dimensions $d_\psi$ and the ghost numbers $gh_\psi$ of
the fields being $\{d_\psi\}=\{1,2,1,1,2,2\}$ and
$\{gh_\psi\}=\{0,0,-1,1,-1,-2\}.$ It is very easy to see that all 1PI
Feynman diagrams with external fields $K^{a\m},H^a$ are superficially
convergent. This ensures that the manipulations performed above to
arrive at the effective action $\gm_{\La M_j m_i}[\psi]$ do not
introduce divergences.

We will write
\begin{displaymath}
  \gm_{\La M_j m_i}[\psi] = \gm^{(0)}_\La\,[\psi]
            + \gm^{(1)}_{\La M_j m_i}[\psi] + O(\hbar^2) \,,
\end{displaymath}
where by definition the tree-level contribution $\gm^{(0)}_\La[\psi]$
is given by
\begin{displaymath}
   \gm^{(0)}_\La[\psi]=S_\La+S_{\rm ext}
\end{displaymath}
and $\gm^{(1)}_{\La M_j m_i}[\psi]$ collects all one-loop corrections.
Since the action $S_\La$ and the determinants $\,\det{\BB A}^1_j$ and
$\,\det{\BB C}^1_i$ are BRS invariant, $\gm_{\La M_j m_i}[\psi]$
satisfies the BRS identities. At tree level and one loop, they take
the form
\begin{eqnarray}
   & \De_\La \,\gm^{(0)}_\La[\psi] = 0  & \label{STzero} \\[9pt]
   & \De_\La\, \gm^{(1)}_{\La M_j m_i}[\psi] = 0 \>, & \label{STone}
\end{eqnarray}
where $\De_\La$ is the Slavnov-Taylor operator for the action
$\gm^{(0)}_\La[\psi].$ The operator $\De_\La$ can be written as
\begin{displaymath}
 \De_\La = \De + \idx {\d S_2\over \d A^a_\m}~{\d\over \d K^{a\m}}~,
\end{displaymath}
where
\begin{equation}
  \De = \idx \bigg[\,
             {\d\gm^{(0)}\over \d A^a_\m}\>{\d\over\d K^{a\m}}
           + {\d\gm^{(0)}\over \d K^{a\m}}\>{\d\over\d A^a_\m}
           + {\d\gm^{(0)}\over \d c^a}\>{\d \over \d H^a}
           + {\d\gm^{(0)}\over \d H^a}\>{\d \over \d c^a}
           + b^a\>{\d\gm^{(0)} \over \d\barc^a} \,\bigg]
\label{STop}
\end{equation}
is the Slavnov-Taylor operator for Yang-Mills theory and $S_2$ is as
in eq. (\ref{HCD2}), the action $\gm^{(0)}[\psi]$ being the
tree-level Yang-Mills action
\begin{equation}
   \gm^{(0)}[\psi] = S_{\YM} + S_{\GF} + S_{\rm ext}\,.
\label{gm0YM}
\end{equation}
For later use, we recall that $\De$ is nilpotent, $\De^2=0,$ and
satisfies $\De S_{\YM}=0.$ Eq. (\ref{STzero}) is trivially satisfied
does not provide any new information. We want to study the structure
of the 1PI Green functions that the regularized effective action
$\gm_{\La M_j m_i}[\psi]$ generates and the constraints that eq.
(\ref{STone}) imposes on them.

Let us then consider the regularized vacuum polarization tensor
$\Pi^{ab}_{\m\n}(p,\La,M_j,m_i).$ Manifest BRS invariance implies
that up to one loop it will be of the form
\begin{equation}
  \Pi^{ab}_{\m\n}(p,\La,M_j,m_i) = \bigg[\, 1 +\, \gtwo \;
     \Pi(p,\La,M_j,m_i)\,\bigg]\,\d^{ab} \,
     \big(\,p^2 g_{\m\n} - p_\m p_\n \,\big) \>,
\label{Piregone}
\end{equation}
where $\Pi(p,\La,M_j,m_i)$ is a dimensionless function of its
arguments. Since $\Pi(p,\La,M_j,m_i)$ is dimensionless, it can be
written without loss of generality as a function of
$p^2/\La^2,~p^2/M_j^2$ and $p^2/m_i^2.$ Power counting implies
that in the limit $\La,M_j,m_i\to\infty,$ $\Pi(p,\La,M_j,m_i)$ will
take the form
\begin{equation}
  \Pi(p,\La,M_j,m_i) \,
      =\, A_3\,\ln\!\bigg( \poLa \bigg)
      +\, B_3 \,\sum_{j=1}^J \,\a_j  \ln\!\bigg( \poMj \bigg)
      +\, C_3 \,\sum_{i=1}^I \,\ga_i \ln\!\bigg( \pomi \bigg)
      +\, \pi_0 \>,
\label{Piregtwo}
\end{equation}
with $A_3,~B_3,~C_3$ and $\pi_0$ coefficients that depend on $\a.$
Assume more generally now that $G^{(0)}(p_e)$ denotes an arbitrary
1PI Yang-Mills Green function at tree level with
$\{p_e\}=\{p_1^\m,\ldots,p_E^\m\}$ independent external momenta.
Manifest BRS invariance and power counting imply that in the limit
$\La,M_j,m_i\to\infty$ the corresponding regularized Green function,
$G(p_e,\La,M_j,m_i),$ can be written up to one loop without loss of
generality as
\begin{displaymath}
   G(p_e,\La,M_j,m_i) = G^{(0)}(p_e) + \gtwo \,\Big[ \,
       \phi_G(p,\La,M_j,m_i) \; G^{(0)}(p_e) \, + \,
       G^{\rm fin} (p_e)\,\Big] \>.
\end{displaymath}
where $\phi_G(p,\La,M_j,m_i)$ is given by
\begin{equation}
\phi_G (p,\La,M_j,m_i) \,=\,
   A_G\,\ln\!\bigg( \poLa\bigg) +\,
   B_G \,\sumj \,\a_j \ln\!\bigg( \poMj\bigg) +\,
   C_G \,\sumi \,\ga_i \ln\!\bigg( \pomi\bigg)
\label{phip}
\end{equation}
and $G^{\rm fin}(p_e)$ collects all finite contributions. The momentum
$p^\m$ in eq. (\ref{phip}) is a non-zero linear combination of the
external momenta and can be chosen arbitrarily since
\begin{displaymath}
   \ln\big(p^2/\La^2\big) = \ln\big(p'^2/\La^2\big)
       + \ln\big(p^2/p'^2\big)
\end{displaymath}
for all $p'^2\neq 0.$ Obviously, different choices of $p^\m$ give
different finite parts $G_{\rm fin}(p_e)$ but the same coefficients
$A_G,~B_G$ and $C_G.$ Introducing a renormalization mass scale $\ka$
and using the Pauli-Villars conditions (\ref{PValpha}) and
(\ref{PVgamma}), we can cast $G(p_e,\La,M_j,m_i)$ as
\begin{equation}
   G(p_e,\La,M_j,m_i) = G^{(0)}(p_e) + \gtwo \,\Big[ \,
       \phi_G(\ka,\La,M_j,m_i) \; G^{(0)}(p_e) \, + \,
       G^{\rm fin}_\ka (p_e)\,\Big] \>,
\label{Green}
\end{equation}
where now
\begin{equation}
\phi_G (\ka,\La,M_j,m_i) \,=\,
   A_G\,\ln\!\bigg( \kaoLa\bigg) +\,
   B_G \,\sumj \,\a_j \ln\!\bigg( \kaoMj\bigg) +\,
   C_G \,\sumi \,\ga_i \ln\!\bigg( \kaomi\bigg)
\label{Greendiv}
\end{equation}
and
\begin{equation}
  G^{\rm fin}_\ka(p_e) = \left( A_G-B_G-C_G \right)\,
           \ln\!\bigg( {p^2\over \ka^2}\bigg) G^{(0)}(p_e)
  + G^{\rm fin}(p_e) \>.
\label{Greenfin}
\end{equation}
In the sequel we will be using the following notation. Subindices
$G=2,3,4$ will respectively refer to the two, three and four-point
Green functions $\VEV{A^2}_{\rm 1PI},\;\VEV{A^3}_{\rm 1PI}$ and
$\,\VEV{\rm A^4}_{\rm 1PI};$ subindices $G=2$ and a tilde and $G=3$
and a tilde will refer to $\VEV{\barc c}_{\rm 1PI}$ and $\,\VEV{\barc
A c}_{\rm 1PI};$ and subindices $G=K$ and $G=H$ will refer to
$\VEV{K\!Ac}_{\rm 1PI}$ and $\VEV{Hcc}_{\rm 1PI}.$ Note that since the
Pauli-Villars fields do not couple to the Faddeev-Popov ghosts nor to
the external fields, we have
\begin{equation}
   \tilde{B}_2 =\tilde{B}_3=0 \qquad \tilde{C}_2 =\tilde{C}_3=0
   \qquad B_K=C_K=0 \qquad B_H=C_H=0\,.
\label{brsghost}
\end{equation}

To find the constraints on the coefficients $A_G,~B_G$ and $C_G$ that
the BRS indentity (\ref{STone}) imposes, we note that one-loop 1PI
diagrams with $K^{a\m}$ external legs have $B_K=C_K=0.$ Hence these
diagrams give contributions to $\gm^{(1)}_{\La M_j m_i}[\psi]$ that
when $\La,M_j,m_i\to\infty$ are either finite or diverge as
$\ln(\ka^2/\La^2).$ Recalling that $S_2$ has a overall factor
$1/\La^4,$ this implies that
\begin{equation}
  \lim_{\La,M_j,m_i\to\infty} ~
      \idx {\d S_2\over \d A^a_\m}~{\d\over \d K^{a\m}}~
      \gm^{(1)}_{\La M_j m_i}[\psi] = 0 \,.
\label{noK}
\end{equation}
{}From our considerations above it follows that in the limit
$\La,M_j,m_i\to\infty$ the one-loop contribution to the regularized
effective action takes the form
\begin{equation}
  \gm^{(1)}_{\La M_j m_i}[\psi]
      = \gm^{\rm (1)\,div}_{\La M_j m_i \ka}[\psi]
      + \gm^{\rm (1)\,fin}_\ka\,[\psi] + \mbox{v.t.}\>,
\label{efflimit}
\end{equation}
where $\gm^{\rm (1)\,div}_{\La M_j m_i \ka}[\psi]$ generates the
$\phi_G\!$-contributions in eq. (\ref{Greendiv}), $\gm^{\rm
(1)\,fin}_\ka\,[\psi]$ generates finite one-loop radiative corrections
and `v.t.' collects contributions that vanish as
$\La,M_j,m_i\to\infty.$ Eqs. (\ref{noK}) and (\ref{efflimit}) imply
that the BRS identity (\ref{STone}) reduces in the limit
$\La,M_j,m_i\to\infty$ to
\begin{eqnarray}
  & \De \gm^{\rm (1)\,div}_{\La M_j m_i \ka}[\psi] = 0
\label{STdiv}\\[9pt]
  & \De \gm^{\rm (1)\,fin}_\ka[\psi] = 0\,. \label{STfin}
\end{eqnarray}
Using that by definition $\gm^{\rm (1)\,div}_{\La M_j m_i
\ka}[\psi]$ generates the $\phi_G\!$-contributions to the Green
functions in eq. (\ref{Greendiv}), it is not difficult to see after
some algebra that eq. (\ref{STdiv}) implies that the coefficients
$A_G,~B_G$ and $C_G$ satisfy
\begin{eqnarray}
  & A_2 - A_3 \,=\, A_3 - A_4 \,= \,
     \tilde{A}_2 - \tilde{A}_3  & \label{brsA} \\
  & B_2 \,=\, B_3 \,=\, B_4 & \label{brsB} \\
  & C_2 \,=\, C_3 \,=\, C_4 \,. & \label{brsC}
\end{eqnarray}
Once we now the form of the divergences in the regularized Green
functions when the regulators are sent to infinity, we are in a
position to renormalize the theory.

To define a one-loop renormalized Yang-Mills theory consistent with
gauge invariance, we have to provide a one-loop renormalized effective
action
\begin{equation}
   \gm_{\WW}\,[\psi] = \gm^{(0)}[\psi]
                   + \gm^{(1)}_{\WW}[\psi] + O(\hbar^2)
\label{effectiveYM}
\end{equation}
that generates finite 1PI Green functions and that satisfies the
Yang-Mills BRS identities. We recall that the latter take the form up
to one loop
\begin{eqnarray}
   & \De\,\gm^{(0)}\,[\psi] = 0  & \label{STYMzero} \\[6pt]
   & \De\, \gm^{(1)}_{\WW}\,[\psi] = 0 \>. & \label{STYMone}
\end{eqnarray}
For $\gm^{(1)}_{\WW}[\psi]$ we take
\begin{equation}
   \gm^{(1)}_{\WW}[\psi] = \gm^{\rm (1)\,fin}_\ka\,[\psi]
            + c_1 S_{\YM} + c_2 \,\De X[\psi]\,,
\label{gm1ren}
\end{equation}
where $\gm^{\rm (1)\,fin}_\ka\,[\psi]$ is the same as in eq.
(\ref{efflimit}), $c_1$ and $c_2$ are arbitrary coefficients of order
$\hbar$ and $X[\psi]$ is the most general local integrated functional
of mass dimension 3 and ghost number $-1$ that can be constructed with
the fields $\psi.$ The coefficients $c_1$ and $c_2$ and the functional
$X$ are taken to be independent of the scale $\ka.$ By construction,
$\gm_{\WW}\,[\psi]$ generates finite Green functions since nothing in
it depends on the cut-offs $\La,~M_j$ and $m_i.$ As regards the BRS
identities, eq. (\ref{STYMzero}) is trivially satisfied, so we only
have to show that eq. (\ref{STYMone}) holds.  That this is the case
follows from eq. (\ref{STone}) and the properties $\De S_{\YM}=0$ and
$\De^2=0.$ Hence the effective action $\gm_{\WW}\,[\psi]$ with
$\gm^{(1)}_{\WW}[\psi]$ as in eq. (\ref{gm1ren}) defines the most
general one-loop renormalized theory compatible with BRS invariance.

Note that we have defined the renormalized theory via an effective
action without making reference to any bare theory. This way to
proceed can be viewed as a Bogoliubov subtraction {\bf T}
on the regularized effective action, namely
\begin{equation}
   \gm_{\WW}\,[\psi] = \lim_{\La,M_j,m_i\to\infty}~
      \left(\,1-\mbox{\bf T}\,\right)\,\gm_{\La M_j m_i}[\psi]\,,
\label{Bogo1}
\end{equation}
where
\begin{equation}
   \mbox{\bf T} \,\gm_{\La M_j m_i}[\psi] =
       \gm^{\rm (1)\,div}_{\La m_j m_i \ka} [\psi]
         - c_1 S_{\YM} - c_2 \,\De X[\psi] \,.
\label{Bogo2}
\end{equation}
The first term on the right-hand side in eq. (\ref{Bogo2}) removes the
divergences in the regularized effective action when the regulators
are sent to infinity. After subtracting this term and taking the limit
$\La,M_j,m_i\to\infty,$ we are left with $\gm^{\rm
(1)\,fin}_\ka[\psi].$ The second and third terms account for arbitrary
finite BRS-invariant local renormalizations. This follows from the
fact \cite{Piguet} that the most general solution to the equation $\De
Y[\psi]=0$ over the space of local integrated functionals of mass
dimension 4 and ghost number zero is precisely of the form $c_1
S_{\YM}+c_2\,\De X[\psi].$ All this implies that the subtraction
$\mbox{\bf T}$ above gives the most general BRS invariant renormalized
effective action that can be defined taking as starting point the
regularized effective action provided by the regularization
prescription we are considering\footnote{For further details on
Bogoliubov-type subtractions for regularization methods containing
higher covariant derivatives see ref. \cite{hcd}}. We would like to
finish this section with two comments.

{\it Comment 1.} The regularized effective action $\gm_{\La M_j
m_i}\,[\psi]$ from which the renormalized effective action
$\gm_{\WW}[\psi]$ has been defined is consistent with a quantum action
principle \cite{Piguet}. Thus the 1PI Green functions generated by
$\gm_{\WW}\,[\psi]$ satisfy the renormalization group equations.

{\it Comment 2.} From the point of view of Yang-Mills theory, the
masses $\La,~M_j$ and $m_i$ are regulators on the same footing and
hence there is no reason why subtraction of divergences should be done
in steps. What the $\mbox{\bf T}$ subtraction in eqs.
(\ref{Bogo1})-(\ref{Bogo2}) does is precisely to subtract the
divergences associated to $\La,~M_j$ and $m_i$ at the same time. If,
say, one performs a first renormalization to subtract the divergences
that occur at finite $\La$ when $M_j$ and $m_i$ are sent to infinity
and then performs a second renormalization to subtract the divergences
that arise when $\La$ goes to infinity, one finds the same of
inconsistencies as in ref. \cite{Seneor}.

\medskip

\mysection{The wrong beta function, a no-go result}

In this section we compute the beta function and the anomalous
dimensions of the fields. To do this, we use the renormalization group
equation.

\mysubsection{The renormalization group equations}

Let us recall that if $G_{\WW}\,(p_e,\ka,g,\a)$ denotes a renormalized
1PI Yang-Mills Green function with $N_\psi$ external legs of type
$\psi,$ the renormalization group equation takes the form
\begin{equation}
   \bigg[ \,\ka~{\partial\over\partial\ka}
        + \b(g)~{\partial\over\partial g}
        + {1\over 2}\,\sum_\psi \ga_\psi(g) \,N_\psi
        + \d(g)~{\partial\over\partial\a}\,\bigg]\,
   G_{\WW}\,(p_e,\ka,g,\a) = 0 \>,
\label{RGeq}
\end{equation}
where the coefficients $\b(g),~\ga_\psi(g)$ and $\d(g)$ are power
series in $g,$
\begin{displaymath}
\begin{array}{l}
  \b(g) = \b_1 \,g + \b_2 \,g^2 + \ldots \\[6pt]
  \ga_\psi (g) = \ga_{\psi,1}\, g + \ga_{\psi,2} \,g^2 + \ldots \\[6pt]
  \d(g) = \d_1\, g + \d_2 \, g^2 +\ldots
\end{array}
\end{displaymath}
The renormalization group equation is an equation in the coefficients
$\b(g),~\ga_\psi(g)$ and $\d(g)$ and holds for all values of the
external momenta $p^\m_e$ and the coupling constant $g.$ We want to
solve this equation up to one loop using as data the renormalized
Green functions generated by the effective action $\gm_{\WW}\,[\psi]$
defined in the previous section.

So let us consider the vacuum polarization tensor generated by
$\gm_{\WW}\,[\psi].$ We remind ourselves that it has the form
\begin{equation}
   \quad\Pi_{\WW\,}{}^{ab}_{\m\n}(p,\ka,g,\a) = - \,\bigg\{ 1 + \gtwo \,
       \bigg[ \, z_2(\a)\, \ln\!\bigg({p^2\over\ka^2}\bigg)
            + \pi_0(\a)\, \bigg] \bigg\} ~\d^{ab}\,
       \Big( p^2 g_{\m\n} - p_\m p_\n \Big) + O(g^4)\>,\quad
\label{Piren}
\end{equation}
where for convenience we have introduced the notation
\begin{equation}
    z_2(\a) = A_2 - B_2 -C_2
\label{z2}
\end{equation}
and $\pi_0(\a)$ is a constant that depends on $\a$ and that already
includes local finite contributions from the terms $c_1S_{\YM}$ and
$c_2\,\De X$ in eq. (\ref{gm1ren}). Substituting
$G_{\WW}\,(p_e,\ka,g,\a)$ in eq. (\ref{RGeq}) by
$\Pi_{\WW\,}{}^{ab}_{\m\n}(p,\ka,g,\a),$ we obtain for the left-hand
side a power series in $g.$ For the equation to hold for all $g,$ the
coefficients of this series must vanish independently. This gives to
orders $g,~g^2$ and $g^3$ the following three equations:
\begin{displaymath}
  \ga_{A,1}=0
\end{displaymath}
\begin{displaymath}
  \ga_{A,2} + \,{\cv\over 16\,\pi^2}\, \bigg[\, 2\,\b_1\,\pi_0(\a)
     - 2\,z_2(\a) + \b_1 \ln\!\bigg({p^2\over\ka^2}\bigg)\bigg] =0
\end{displaymath}
\begin{displaymath}
   \ga_{A,3} + \,{\cv\over 16\,\pi^2}\,\bigg\{ 2\,\b_2\,\pi_0(\a)
     + \ga_{A,1} + \d_1\,{\partial\pi_0(\a)\over\partial\a}
     + \bigg[ \Big( 2\,\b_2 + \ga_{A,1}\Big)\, z_2(\a)
            + \d_1\,{\partial z_2(\a)\over\partial\a}\,\bigg]
            \ln\!\bigg({p^2\over\ka^2}\bigg) \bigg\} =0 \,.\\[12pt]
\end{displaymath}
Since the renormalization group equation holds for all values of
$p^\m,$ terms independent of $p^\m$ and terms proportional to
$\ln(p^2/\ka^2)$ must vanish separately. This implies in particular
\begin{displaymath}
\begin{array}{cc}
  \ga_{A,1} = 0 \qquad & {\ds
  \ga_{A,2} = {\cv\over 8\pi^2}~z_2(\a) }\\[12pt]
  \b_1=0 \qquad & {\ds
  \b_2 = {1\over 2\,z_2(\a)}~{\partial z_2(\a)\over\partial\a}~\d_1~.}
\end{array}
\end{displaymath}
Let us next take the renormalized ghost self-energy. Our arguments in
the previous section implies that it has the form
\begin{equation}
   \Om_{\WW}^{ab}(p,\ka,g,\a) = \bigg\{ 1 + \gtwo \,
       \bigg[ \, \tilde{z}_2(\a)\, \ln\!\bigg({p^2\over\ka^2}\bigg)
            + \om_0(\a)\, \bigg] \bigg\} ~\d^{ab}\,p^2 + \,O(g^4)\>,
\label{ghostren}
\end{equation}
where $\tilde{z}_2(\a)$ stands for
\begin{equation}
  \tilde{z}_2(\a) = \tilde{A}_2
\label{z2tilde}
\end{equation}
and $\om(\a)$ is an $\a\!$-dependent constant. Proceeding in the same
way as for the vacuum polarization tensor, it is straightforward to
see that the renormalization group equation for the ghost self-energy
yields
\begin{displaymath}
\begin{array}{cc}
  \ga_{c,1} + \ga_{\barc,1} = 0 \qquad & {\ds
  \ga_{c,2} + \ga_{\barc,2} = {\cv\over 4\pi^2}~\tilde{z}_2(\a)
  }\\[12pt]
  \b_1=0 \qquad & {\ds
  \b_2 = {1\over 2\,\tilde{z}_2(\a)}
          ~ {\partial \tilde{z}_2(\a)\over\partial\a}~\d_1 ~. }
\end{array}
\end{displaymath}
Let us finally look at the renormalization group equation for the
$\barc Ac\!$-vertex. If $k^\m$ and $p^\m$ denote the momenta of the
incoming ghost and the outgoing antighost, the vertex can be written
without loss of generality as
\begin{equation}
   V_{\WW}{}_\m^{abc\,}(p,k,\ka,g,\a) = -\,ig\,f^{abc} \,\bigg\{
      \bigg[\, 1 + \gtwo ~\tilde{z}_3(\a)\,
                   \ln\!\bigg({p^2\over\ka^2}\bigg) \bigg] p_\m
      + \gtwo~ V^{\rm fin}_\m(k,p,\a) \bigg\} +\,O(g^5) \>,
\label{vertexren}
\end{equation}
where the coefficient $\tilde{z}_3(\a)$ is given by
\begin{equation}
  \tilde{z}_3(\a) = \tilde{A}_3
\label{z3}
\end{equation}
and $V^{\rm fin}_\m(k,p,\a)$ conatins local as well as non-local
finite $\ka\!$-independent radiative corrections. Using eq.
(\ref{vertexren}) as input in the renormalization group equation, eq.
(\ref{RGeq}), expanding in powers of $g$ up to order two and using our
previous results for $\b_1$ and the anomalous dimensions
$\ga_{\psi,1}$ and $\ga_{\psi,2},$ it is very to see that
\begin{displaymath}
   \b_2=0 \qquad
   \b_3 = {\cv\over 16\,\pi^2}~\Big[\, 2\,\tilde{z}_3(\a) -
            2\,\tilde{z}_2(\a) - z_2(\a)\,\Big]\,.
\end{displaymath}
The first one of these equation implies in turn that
\begin{displaymath}
  \d_1=0\,.
\end{displaymath}
Thus the renormalization group coefficients to lowest order can all
be expressed in terms of $z_2(\a),~\tilde{z}_2(\a)$ and
$\tilde{z}_3(\a).$ To actually compute $z_2(\a),~\tilde{z}_2(\a)$ and
$\tilde{z}_3(\a),$ we must explicitly compute the limit
$\La,M_j,m_i\to\infty$ of the regularized vacuum polarization tensor,
the regularized ghost self-energy and the regularized $\barc A
c\!$-vertex. We do this in the sequel.

\mysubsection{Explicit computations}

We begin with the vacuum polarization tensor $\Pi^{ab}_{\m\n}
(p,\La,M_j,m_i).$ At one loop, it receives contributions from the
Feynman diagrams in Figs. 1 and 2 and in the limit
$\La,M_j,m_i\to\infty$ it takes the form in eqs. (\ref{Piregone}) and
(\ref{Piregtwo}). A look at the Feynman rules in Appendix A shows that
\begin{list} {}{\setlength{\leftmargin}{40pt}}
\vspace{-18pt}
\item[(i)] diagrams (1e) and (2b) do not depend on any mass at all,
\vspace{-9pt}
\item[(ii)] diagrams (1a) and (1c) depend on $\La,$
\vspace{-9pt}
\item[(iii)] diagrams (1f) and (2e) depend on $m_i,$ and
\vspace{-9pt}
\item[(iv)] diagrams (1b), (1d), (2a), (2c) and (2d) depend on both
$\La$ and $M_j.$
\vspace{-18pt}
\end{list}
This implies that the coefficient $C_2$ in eq. (\ref{Piregtwo}) is
unambiguous, whereas the coefficients $A_2$ and $B_2$ depend on the
path in $(\La,M_j)\!$-space followed to approach $\La,M_j=\infty.$ Our
calculations in Appendix B yield for $C_2$ the value
\begin{equation}
   C_2 = -\, {1\over 3} ~.
\label{Clarge}
\end{equation}
As regards $A_2$ and $B_2,$ we consider in this paper paths of type
\begin{equation}
  \La,M_j\to\infty \quad {\rm with} \quad  {M_j\over\La} = \th_j
  \quad {\rm and} \quad 0 \leq \th_j \leq \infty \,.
\label{path}
\end{equation}
The path $\th_j=\infty$ corresponds to first taking the limit
$M_j\to\infty$ at finite $\La,$ and then taking the limit
$\La\to\infty.$ The path $\th_j=0$ on the contrary corresponds to
first taking the limit $\La\to\infty$ at finite $M_j,$ and then taking
the limit $M_j\to\infty.$ The calculations in Appendix B show that
\begin{eqnarray}
  \th_j=\infty: & ~{\ds A_2 = 6- {\a\over 2} }
                & ~B_2 = 4  \label{PiLM} \\[12pt]
  \th_j=0: & ~{\ds A_2 = -\,{3+\a\over 2} }
           & ~{\ds B_2 = -\,{21\over 6} } \label{PiML} \\[12pt]
  0<\th_j<\infty: & ~{\ds A_2 = 2 -{\a\over 2} }
                  & ~B_2=0 \,. \label{PiL}
\end{eqnarray}
We see that even though $A_2$ and $B_2$ are different for different
$\th_j,$ the combination $A_2-B_2$ is the same. These results yield
for $z_2(\a)$ in eq. (\ref{z2})
\begin{equation}
   z_2(\a)= {7\over 3} - {\a\over 2} ~ .
\label{z2result}
\end{equation}
It is important to note that in general the coefficients $A_G,~B_G$
and $C_G$ are not important by themselves. What is really important
are the combinations $z_G(\a)=A_G-B_G-C_G,$ since they carry the
$\ka\!$-dependence of the renormalized Green functions. Note as a
matter of fact that the beta function depends precisely on these
combinations. In a sense, we can think of every $(\La,M_j)\!$-path,
equivalently of every set $\{A_G,B_G,C_G\},$ as defining a different
regularization method for which the BRS identities
(\ref{brsA})-(\ref{brsC}) hold. Since we will extensively refer later
on to the renormalized vacuum polarization tensor, let us put eqs.
(\ref{Piren}) and (\ref{z2result}) together:
\begin{equation}
  \Pi_{\WW}{}^{ab}_{\m\n}\,(p,\ka,g,\a) = -\,\bigg\{ 1 + \gtwo
       \, \bigg[\, \bigg( {7\over 3} -{\a\over 2} \bigg)
       \ln\!\bigg({p^2\over \ka^2}\bigg) + \pi_0(\a)\, \bigg]
  \bigg\} \,\d^{ab} \Big( p^2 g_{\m\n} - p_\m p_\n \Big) \,.
\label{Piwrong}
\end{equation}

Next we move on to the ghost self-energy and the $\barc Ac\!$-vertex.
The regularized ghost self-energy at one loop $\Om^{ab}(p,\La)$
receives contributions from only the diagram in Fig. 3. This diagram
is finite by power counting and, apart from the external momentum
$p^\m,$ only depends on the mass $\La.$ To evaluate its limit
$\La\to\infty,$ we use the same techniques as for the vacuum
polarization tensor and obtain
\begin{displaymath}
   \Om^{{\rm (1)} \,ab}(p,\La) = \gtwo\,\bigg[\,{3-\a\over 4}\>
   \ln\!\bigg(\poLa\bigg) + \om_0(\a) \,\bigg] \,\d^{ab} \,p^2 \,.
\end{displaymath}
This gives for $\tilde{z}_2(\a)$ in eqs.
(\ref{ghostren})-(\ref{z2tilde}) the value
\begin{equation}
   \tilde{z}_2(\a) = \,{3-\a\over 4} ~.
\label{tildez2result}
\end{equation}
As concerns the $\barc A c\!$-vertex, at one loop it receives
contributions from the two diagrams in Fig. 4. These diagrams are
finite by power counting and depend on the external momenta $k^\m$ and
$p^\m$ and the mass $\La.$ Using the same techniques as in Appendix B,
it is not difficult to see that
\begin{displaymath}
   V^{{\rm (1)}\,abc}_\m(k,p,\La) = -\,ig\,f^{abc}\,\gtwo\,\bigg[
       -\,{\a\over 2}\> \ln\!\bigg(\poLa\bigg) \,p_\m
       + \,\a\,V_\m^{\rm fin}(k,p,\a)\,\bigg]
\end{displaymath}
as $\La$ goes to infinity. This yields
\begin{equation}
   \tilde{z}_3 = -\,{\a\over 2}~.
\label{z3result}
\end{equation}

Substituting the values obtained for $z_2(\a),~\tilde{z}_2(\a)$ and
$\tilde{z}_3(\a)$ in the expressions for the beta functions and the
anomalous dimensions above, we have
\begin{eqnarray}
  & {\ds \qquad \b(g) = -\>{23 \over 6}\>{g^3\cv\over 16\pi^2}
    + O(g^4)} \qquad &
    \ga_A(g) = \bigg( {14\over 3} - \a\bigg)\, \gtwo + O(g^3)
  \label{wrong}\\[12pt]
  & {\ds \ga_c + \ga_{\barc} = (3-\a)~\gtwo + O(g^3)} &
    \qquad\qquad \quad \d(g) = O(g^2) \,. \nonumber
\end{eqnarray}
The results for the beta function and the anomalous dimension of the
gauge field are in disagreement with the well known results
\begin{equation}
  \b(g) = -\>{11 \over 3}\>{g^3\cv\over 16\,\pi^2} + O(g^4)
  \qquad
  \ga_A(g) = \bigg( {13\over 3} -\a \bigg)\, \gtwo + O(g^3) \,.
\label{correct}
\end{equation}
We recall that for unitary theories the one and two-loop coefficients
in the beta function and the one-loop coefficient of the anomalous
dimensions are universal quantities that do not depend on the
regularization method nor renormalization scheme used to compute them.
In view of this, we have to conclude that the regularization
prescription under consideration does not yield a consistent
renormalized Yang-Mills theory, hence that it must be discarded as a
candidate to regularize non-abelian gauge theories.

\medskip

\mysection{Non-physical radiative corrections from Pauli-Villars
determinants}

The beta function at one loop is build up with radiative corrections
of type $\ln(p^2/\ka^2).$ Hence, the fact that the result in eq.
(\ref{wrong}) is not correct indicates that the regularization method
generates somewhere non-local logarithmic radiative corrections which
should not be there. To find these corrections and their origin, we
proceed as follows.

Suppose that instead of the regularization prescription of Faddeev
and Slavnov we use dimensional regularization or a plain cut-off.
Then the vacuum polarization tensor, the ghost self-energy and the
ghost-vertex take after renormalization the form
\begin{equation}
  \Pi'_{\WW}{}^{ab}_{\m\n}\,(p,\ka,g,\a) = -\,\bigg\{ 1
     + \gtwo \, \bigg[\, \bigg( {13\over 6} -{\a\over 2} \bigg)
            \ln\!\bigg({p^2\over\ka^2}\bigg) + \pi_0(\a)\,
      \bigg] \bigg\} \,\d^{ab} \Big( p^2 g_{\m\n}- p_\m p_\n \Big)
\label{Picorrect}
\end{equation}
\begin{eqnarray}
  &{\ds \Om'^{ab}_{\WW}\,(p,\ka,g,\a) = \bigg\{ 1 + \gtwo
      \bigg[ \, {3-\a\over 4}\,\ln\!\bigg({p^2\over \ka^2}\bigg)
       + \om_0(\a)\, \bigg] \bigg\} \d^{ab} \,p^2 }  & \nonumber
    \\[12pt]
  &{\ds V'_{\WW}{}\m^{abc} \,(k,p,\ka,g,\a) =-\,ig\,f^{abc} \bigg\{
    \bigg[\,1 - \gtwo \> {\a\over 2}\,
           \ln\!\bigg({p^2\over \ka^2}\bigg)\bigg] \, p_\m
       + \gtwo ~\a \, V_\m^{\rm fin}(k,p,\a)\,\bigg\}\>. } &
    \nonumber
\end{eqnarray}
{}From general theorems of renormalization theory, we know that two
renormalized series for a 1PI Green function that agree up to order
$\hbar^{n-1}$ can differ at order $\hbar^n$ by at most a local term.
It happens however that the renormalized series for the vacuum
polarization tensor in eqs. (\ref{Piwrong}) and (\ref{Picorrect})
agree at tree level but differ at one loop by a non-local term. The
question that arises then is whether eqs. (\ref{Piwrong}) and
(\ref{Picorrect}) correspond to different but consistent quantum
Yang-Mills theories. To answer this question, we remind that in
Minkowski space the renormalized vacuum polarization tensor takes up
to one loop the form
\begin{displaymath}
    \Pi_{\MM}{}_{\m\n}^{ab}(p,\ka,g,\a) = -\,\bigg\{ 1 + \gtwo \,
      \bigg[\, c(\a) \,\ln\!\bigg( {-\,p^2\!+i0^+\over \ka^2} \bigg)
      + \pi_0(\a)\,\bigg]\bigg\} \,\d^{ab}\,
      \Big(p^2 g_{\m\n} - p_\m p_\n\Big) \>.
\end{displaymath}
The coefficient $c(\a)$ is uniquely determined by the imaginary
part of $\Pi_{\MM}{}_{\m\n}^{ab},$
\begin{equation}
   {\rm Im}\> \Pi_{\MM}{}_{\m\n}^{ab}(p,\ka,g,\a) = -\,c(\a)~\gtwo~
   \th(p^2)\, \left( p^2 g_{\m\n} - p_\m p_\n \right) \>,
\label{Cutkosky}
\end{equation}
But the imaginary part of $\Pi_{\MM}{}_{\m\n}^{ab}$ can be computed by
means of Cutkosky rules as
\begin{center}\begin{picture}(305,80)(0,0)
\Text(80,40)[r]{${\rm Im} \left(\Pi_{\MM}{}^{ab}_{\m\n}\right)$}
\Text(90,40)[c]{$=$}
\Photon(105,40)(125,40){2}{2.5}
\PhotonArc(145,40)(20,0,180){2}{6} \Vertex(125,40){1.5}
\PhotonArc(145,40)(20,180,360){2}{6} \Vertex(165,40){1.5}
\Photon(166,40)(185,40){2}{2.5}
\Text(200,40)[c]{$+$}
\Photon(215,40)(235,40){2}{2.5}
\DashArrowArc(255,40)(20,0,180){3} \Vertex(235,40){1.5}
\DashArrowArc(255,40)(20,180,360){3} \Vertex(275,40){1.5}
\Photon(275,40)(295,40){2}{2.5} \Text(305,40)[c]{,}
\Line(130,5)(160,75)
\Line(132.5,10.83)(137.5,10.83)
\Line(135,16.66)(140,16.66)
\Line(137.5,22.5)(142.5,22.5)
\Line(140,28.33)(145,28.33)
\Line(142.5,34.16)(147.5,34.16)
\Line(145,40)(150,40)
\Line(147.5,45.83)(152.5,45.83)
\Line(150,51.66)(155,51.66)
\Line(152.5,57.49)(157.5,57.49)
\Line(155,63.33)(160,63.33)
\Line(157.5,69.16)(162.5,69.16)
\Line(240,5)(270,75)
\Line(242.5,10.83)(247.5,10.83)
\Line(245,16.66)(250,16.66)
\Line(247.5,22.5)(252.5,22.5)
\Line(250,28.33)(255,28.33)
\Line(252.5,34.16)(257.5,34.16)
\Line(255,40)(260,40)
\Line(257.5,45.83)(262.5,45.83)
\Line(260,51.66)(265,51.66)
\Line(262.5,57.49)(267.5,57.49)
\Line(265,63.33)(270,63.33)
\Line(267.5,69.16)(272.5,69.16)
\end{picture}\end{center}
where the cut propagators and the vertices are the cut propagators and
vertices of unregularized Yang-Mills theory. The cut-diagrams on the
right-hand side are very easy to calculate in the Feynman gauge. Doing
so and using eq. (\ref{Cutkosky}), we get $c(1)=5/3.$ This selects
the vacuum polarization tensor in eq. (\ref{Picorrect}) and disregards
that in eq. (\ref{Piwrong}). It also shows that the vacuum
polarization tensor as computed with higher covariant derivatives and
Pauli-Villars fields contains unphysical logarithmic radiative
corrections.

To pin-point the origin of these unphysical radiatives corrections, we
consider all the diagrams that contribute to the vacuum polarization
tensor in dimensional regularization and compute their limit $D\to 4.$
Two types of contributions arise: contributions finite at $D=4$, and
contributions singular at $D=4.$ In the first ones, we set $D=4$ and
take the limit $\La,M_j,m_i\to\infty.$ In the second ones, we leave
$D$ arbitrary and keep the masses $\La,~M_j$ and $m_i$ finite.
Proceeding in this way we obtain the results given in Appendix B. It
follows from them that only diagrams (1a), (1e), (2a) and (2b) give
$\ln(p^2)\!$-contributions. The $\ln(p^2)\!$-contribution from
diagrams (1a) and (1e) can be read off from eqs. (\ref{1a}) and
(\ref{1e}):
\begin{equation}
   {\rm (1a)+(1e)} ~~ \to ~~
     - \, \gtwo\>\d^{ab}\, \bigg({13\over 6}-{\a\over 2}\bigg)
     \ln(p^2)\,\left( p^2 g_{\m\n} - p_\m p_\n \right)\>.
\label{Im1a1e}
\end{equation}
As for diagrams (2a) and (2b), their $\ln(p^2)\!$-contribution is [see
eqs. (\ref{2b}), (\ref{2aLM}), (\ref{2aML}) and (\ref{2aL})]
\begin{equation}
   {\rm (2a)+(2b)} ~~ \to ~~  - \, \gtwo\>\d^{ab}\> {1\over 6}\>
     \ln(p^2)\,\left( p^2 g_{\m\n} - p_\m p_\n \right)\>.
\label{Im2a2b}
\end{equation}
It thus becomes clear that diagrams (1a) and (1e) give by themselves
the correct non-local part of the vacuum polarization tensor, whereas
diagrams (2a) and (2b) are responsible for the unwanted logarithmic
corrections. The problem is that without diagrams (2a) and
(2b) the regularized vacuum polarization tensor is not finite nor
transverse. This, together with the observation that diagrams (2a)
and (2b) are generated by the determinants $\,\det{\BB A}^1_j,$ leads
to the conclusion that the very same Pauli-Villars determinants that
regularize divergences also generate unphysical non-local logarithmic
radiative corrections.

The determinant $\,\det{\BB A}^1_j$ has a mass $\La$ hidden in it.
One would like to know if this dependence on $\La$ plays any part in
the generation of the unphysical logarithmic radiative corrections. To
see if this is the case, we replace the action $S_\La$ in the
definition of $\,\det{\BB A}^1_j$ with the Yang-Mills gauge-fixed
action $S$ in eq. (\ref{action}) and re-compute the vacuum
polarization tensor. The Feynman rules are now different and can be
obtained from those in Appendix A by simply setting $\La$ to infinity.
The diagrams that make the vacuum polarization tensor at one loop are
the same as in Figs. 1 and 2, with the difference that now
they are constructed with different Feynman rules. To be on the safe
side, and since Pauli-Villars determinants may not prove sufficient to
regularize all UV divergences, we introduce in addition dimensional
regularization.  Using eqs. (\ref{PValpha}) and (\ref{PVgamma}), and
after some lengthy but straightforward calculations, we obtain
\begin{displaymath}
  \Pi^{(1)\,ab}_{\m\n}(p,M_j,m_i,\ee,\m) = -\,\gtwo\> \d^{ab}\>
     \Pi(p,M_j,m_i) \> \big( p^2 g_{\m\n} - p_\m p_\n \big) ~,
\end{displaymath}
where $ \Pi(p,M_j,m_i,\ee,\m)$ is given by
\begin{displaymath}
   \Pi(p,M_j,m_i,\ee,\m) = -\,{3\!+\!\a\over 2} \> L(p^2)
   - {7\over 2}\, \sumj \a_j \ln\!\bigg(\poMj\bigg)
   - {1\over 3}\, \sumi \ga_i \ln\!\bigg(\pomi\bigg)\!
   + \pi_0(\a,M_j,m_i)\>.
\end{displaymath}
Here $L(z^2)$ denotes
\begin{equation}
  L(z^2)= {1\over \ee} + \ln\!\bigg({z^2\over\m^2}\bigg)
  \qquad\quad D=4+2\ee
\label{Lz}
\end{equation}
and $\pi_0(\a,M_j,m_i)$ collects local contributions which are finite
at $D=4$ and either vanish or remain finite as $M_j,m_i\to\infty.$ The
coefficient in front of $\ln(p^2)$ is ${\ds {7\over 3}-{\a\over 2}},$
the same as in eqs. (\ref{Piren}) and (\ref{z2result}). Furthemore,
although we have not given here partial results, it happens again that
only diagrams (1a), (1e), (2a) and (2b) generate
$\ln(p^2)\!$-contributions and that the latter contributions have the
same form as in eqs. (\ref{Im1a1e}) and (\ref{Im2a2b}). Hence the
origin of the unphysical radiative corrections is not the $\La$ hidden
in $S_\La$ in but the structure of the determinants $\,\det{\BB
A}^1_j.$

Let us further argue that the non-local logarithmic contribution
arising from diagrams (2a) and (2b) can not be set to zero by sending
the masses $\La$ and $M_j$ to infinity in a clever way. Suppose we
take the diagrams in Figs. 1 and 2 and proceed as follows:
\begin{list} {} {\setlength{\leftmargin}{40pt}}
\vspace{-18pt}
\item[(i)] Set the masses $\La,~M_j$ and $m_i$ to infinity in the
Feynman integrands and keep only those diagrams whose integrands do
not vanish.
\vspace{-9pt}
\item[(ii)] Wick rotate to Minkowski's momentum space the
surviving diagrams.
\vspace{-9pt}
\item[(iii)] Compute the imaginary part of the analytically continued
diagrams by means of Cutkosky rules (for a general reference on cut
diagrams see \cite{Diagrammar}).
\vspace{-18pt}
\end{list}
Then, it is not difficult to see that only diagrams (1a), (1e), (2a)
and (2b) develop an imaginary part and that in the Feynman gauge the
latter takes the form
\begin{displaymath}
\begin{array}{l} {\ds
   {\rm Im}\> \Big[ \, {\rm (1a) +(1e)}\,\Big]
    = -\,{5\over 3}\> \gim~ \th(p^2)\, \left( p^2 g_{\m\n}
    - p_\m p_\n \right) }\\[12pt] {\ds
   {\rm Im}\> \Big[ \, {\rm (2a) +(2b)}\,\Big]
    = -\,{1\over 6}\> \gim~ \th(p^2)\, \left( p^2 g_{\m\n} - p_\m p_\n
\right) \>. }
\end{array}
\end{displaymath}
So diagrams (1a) and (1e) keep giving physical logarithmic radiative
corrections and diagrams (2a) and (2b) keep giving unphysical
logarithmic radiative corrections. The formal manipulations in steps
(i)-(iii) are the standard manipulations used to compute the imaginary
part of vacuum polarization tensor in QED \cite{Lifshitz} and rely on
the following observation. In Minkowski space only diagrams with the
topology of diagram (1a) develop an imaginary part at one loop. When
the regulators are switched off, these diagrams are given by integrals
of the form
\begin{displaymath}
  \idq  {P_{\m\n}(q,p)\over (q^2+i0^+)\> [(q+p)^2 +i0^+]} ~,
\end{displaymath}
where $P_{\m\n}(q,p)$ is some polynomial in the momenta $q^\m$ and
$p^\m.$ Rigorously speaking, this integral is a distribution which is
not well defined.  The point is that it can be properly defined
without touching its imaginary part; this is what regularization and
renormalization is all about. Therefore, the imaginary part of the
integral before and after regularization and renormalization should be
the same. Note, as a matter of fact, that for the $(\La,M_j)\!$-paths
considered in this paper, this simple way to proceed gives after Wick
rotation to Euclidean space the same results as the explicit
computations of Appendix B. It thus seems that the process in steps
(i)-(iii) is correct if one is interested only in the imaginary part
of the vacuum polarization tensor in Minkowski space, equivalently the
non-local part in Euclidean space. Furthermore, since steps (i)-(iii)
do not depend on the path in $(\La,M_j)\!$-space followed to send
$\La$ and $M_j$ to infinity, one would expect the renormalized vacuum
polarization tensor to have the form in eqs. (\ref{Piren}) and
(\ref{z2result}) for all $(\La,M_j)\!$-paths

The question that arises now is whether it is possible to amend the
regularization method so as to get rid of the unphysical non-local
radiative corrections. This is possible but at the cost of losing
manifest BRS invariance. To see this, let us assume that we take the
gauge non-invariant version $r=0$ of the Pauli-Villars determinants,
and let us take for concreteness the $(\La,M_j)\!$-path in eq.
(\ref{path}) with $\th_j=\infty.$ Now there are only six Feynman
diagrams that contribute to the vacuum polarization tensor, those in
Fig. 1. Evaluating them in dimensional regularization and taking the
limit $\La,M_j,m_i\to\infty$ in those contributions which are finite
at $D=4,$ we obtain (see Appendix B for partial results)
\begin{equation}
\begin{array}{l} {\ds
  \Pi^{(1)\,ab}_{\m\n}(p,\La,M_j,m_i)\bigg\vert_{r=0} \! =
    \gtwo \>\d^{ab}  \, \bigg\{ \!
          - {\pi\sqrt{3}\over 4} \sumj \a_j\, M_j^{2/3}
                                              \La^{4/3}\,g_{\m\n}
          - {1\over 2} \sumi \ga_i \ln\!\bigg({m_i^2\over\m^2}\bigg)
                                       \,m_i^2 \,g_{\m\n} }\\[12pt]
\phantom{\hspace{30pt}} {\ds
   - \bigg[\, \bigg( {31\over 18} +{\a\over 2} \bigg)
               \ln\!\bigg(\poLa\bigg)
             + {143\over 36} \sumi \a_j \ln\!\bigg(\poMj\bigg)
             - {1\over 12} \sumi \ga_i \ln\!\bigg(\pomi\bigg)
             + \const\, \bigg]\, p^2 g_{\m\n} }\\[12pt]
\phantom{\hspace{30pt}} {\ds
   + \bigg[\,\bigg( {17\over 9} +{\a\over 2} \bigg)
               \ln\!\bigg(\poLa\bigg)
             + {35\over 9} \sumi \a_j \ln\!\bigg(\poMj\bigg)
             + {1\over 6} \sumi \ga_i \ln\!\bigg(\pomi\bigg)
             + \const\, \bigg]\, p_\m p_\n \bigg\} \,.}
\end{array}
\label{r0PiLM}
\end{equation}
Here we have imposed conditions (\ref{PValpha}), (\ref{PVgamma}) and
(\ref{PVmass}), which we recall from subsection 2.2.2 ensure
regularization of divergences. The right-hand side in eq.
(\ref{r0PiLM}) is not transverse, hence BRS invariance does not hold
at the regularized level. To end up with a transverse renormalized
vacuum polarization tensor, we proceed as follows. We further restrict
the Pauli-Villars masses and parameters to satisfy
\begin{equation}
   \sumj \a_j M_j^{2/3} = 0  \qquad\quad
   \sumi \ga_i \,m_i^2 \,\ln(m_i^2) =0 \,.
\label{r0LM}
\end{equation}
By doing this we get rid of the first two terms in eq. (\ref{r0PiLM}).
Next we introduce a renormalization mass scale $\ka$ and perform a
suitable subtraction. The fact that the coefficient in front of
$\ln(p^2)$ in both the terms $p^2g_{\m\n}$ and $p_\m p_\n$ in eq.
(\ref{r0PiLM}) is the same and equal to ${\ds {13\over 6}-{\a\over
2}}$ --compare with eq. (\ref{Picorrect})-- ensures that the
renormalized vaccum polarization tensor will be transverse and will
have the correct non-local part. If instead of $(\La,M_j)\!$-paths
with $\th_j=\infty$ we take paths with $0\leq\th_j < \infty$ and use
the results in Appendix B, it is very easy to see that the same
conclusion holds. The only difference is that to ensure
transversality, we must replace the condition on the masses $M_j$ in
eq. (\ref{r0LM}) with
\begin{displaymath}
\sumj \, \a_j\,M_j^2 = 0 \>.
\end{displaymath}
It thus becomes clear that the terms introduced when $r=0$ is replaced
with $r=1$ so as to have BRS invariance at the regularized level
originate the unphysical non-local logarithmic corrections. Note in
this regard that diagrams (2a) and (2b), the diagrams that originated
the unphysical contributions, do not exist at all for $r=0.$ The
problem with taking $r=0$ is that manifest BRS is lost, thus becoming
uninteresting.

\medskip

\mysection{Conclusion}

In this paper we have addressed the question of whether higher
covariant derivative Pauli-Villars regularization as proposed in refs.
\cite{Slavnov} and \cite{Faddeev} leads to a consistent renormalized
Yang-Mills theory. The regularization prescription is described by the
generating functional $Z_1[J,\chi,\eta,\bareta]$ in eq.
(\ref{functional}) and combines a higher covariant derivative term in
the action with a set of Pauli-Villars determinants $\,\det{\BB
A}^1_j$ and $\,\det{\BB C}^1_i.$ The main results of our investigation
can be summarized in the following three points:

(1) There are certain subtleties regarding the regularization
mechanism that have gone unnoticed in the previous literature and that
reveal that, even after imposing suitable Pauli-Villars conditions,
the regularized Green functions are not finite in the power counting
sense advertized in ref. \cite{Faddeev}. This has been explicitly
realized in subsection 2.2.2, where it is shown that checking that the
regularized vacuum polarization tensor is finite for finite values of
the regulators requires introducing a further regulator $\cR.$ This
gives lie to the claim \cite{Slavnov} \cite{Faddeev} that the
functional $Z_1[J,\chi,\eta,\bareta]$ constitutes by itself a
regularized expression for the Yang-Mills path integral. Yet, since
the non-local part of the regularized Green functions is the same for
all admissible regulators $\cR,$ one can think of using the functional
$Z_1[J,\chi,\eta,\bareta]$ to understand physical properties like \eg\
unitarity.

(2) The regularization prescription does lead to a renormalized theory
consistent with gauge invariance. This settles down some controversy
in the previous literature \cite{Seneor} \cite{Day}.

(3) The Pauli-Villars determinants on which the regularization method
is based generate unphysical non-local logarithmic radiative
corrections. These corrections, being non-local, survive
renormalization and give an inconsistent renormalized Yang-Mills
theory. In particular, the beta function of the resulting renormalized
theory has a one-loop coefficient equal to $-23/6,$ in contradiction
with general results from renormalization theory that state that it
should be $-11/3.$ We expect these logarithmic unphysical corrections
to break unitarity in Minkowski spacetime when the theory is coupled
to matter. In fact, if unitarity is formulated in the sense of van Dam
and Veltman \cite{vanDam}, our discussion in section 5 already shows
that this is the case. We have also seen that it is possible to modify
the Pauli-Villars determinants so as to get rid of the unwanted
radiative corrections, but this is at the expense of losing gauge
invariance at the regularized level.

In view of all this we have to conclude that higher covariant
derivative Pauli-Villars regulators do not provide an admissible gauge
invariant regularization method.
\vspace{24pt}

\noindent{\large\bf Acknowledgements}
\vspace{8pt}

\noindent
The authors are grateful to G. 't
Hooft, J.C. Taylor and M. Veltman for discussions. FRR was supported
by FOM, The Netherlands. Partial support from CICyT, Spain is also
acknowledged.

\vspace{24pt}

\noindent{\large\bf Appendix A:~~ Feynman rules} \vspace{8pt}
{\renewcommand{\theequation}{A.\arabic{equation}}\setcounter{equation}{0}
\setcounter{section}{0}

\noindent
Here we list the Feynman rules corresponding to the functional
$Z_r[J,\chi,\eta,\bareta]$ in eq. (\ref{functional}). The propagators
are given by
\begin{center}\begin{picture}(330,40)(0,0)
\Photon(10,20)(70,20){3}{6}\Vertex(10,20){2}\Vertex(70,20){2}
\Text(10,12)[tc]{$A^a_\m$} \Text(40,30)[bc]{${\ds p\atop \to}$}
\Text(70,12)[tc]{$A^b_\n$}
\Text(90,20)[c]{$=$}
\Text(100,20)[l]{$ \ds
    {\d^{ab}\La^4 \over p^4\,(p^4+\La^4)} \>
    \bigg[ \, p^2 g_{\m\n} -\; p_\m p_\n \;\bigg( 1
                  - {\a\,\La^4\over p^4+\La^4} \bigg)\,\bigg] $}
\end{picture}\end{center}
\begin{center}\begin{picture}(330,40)(0,0)
\Line(10,20)(40,20)\Photon(40,20)(70,20){3}{3}
\Vertex(10,20){2}\Vertex(70,20){2}
\Text(10,12)[tc]{$b^a$} \Text(40,30)[bc]{${\ds p\atop \to}$}
\Text(70,12)[tc]{$A^b_\m$}
\Text(90,20)[c]{$=$}
\Text(100,20)[l]{$ \ds -\,i\,\d^{ab}~{p_\m \over p^2} $}
\end{picture}\end{center}
\begin{center}\begin{picture}(330,40)(0,0)
\DashArrowLine(10,20)(70,20){6}
\Vertex(10,20){2}\Vertex(70,20){2}
\Text(10,12)[tc]{$\barc^a$} \Text(40,30)[bc]{$p$}
\Text(70,12)[tc]{$c^b$}
\Text(90,20)[c]{$=$}
\Text(100,20)[l]{$ \ds -\, {\d^{ab} \over p^2} $}
\end{picture}\end{center}
\begin{center}\begin{picture}(330,40)(0,0)
\Gluon(10,20)(70,20){4}{6}\Vertex(10,20){2}\Vertex(70,20){2}
\Text(10,12)[tc]{$A^{~\,a}_{j\,\m}$}
\Text(40,30)[bc]{${\ds p\atop \to}$}
\Text(70,12)[tc]{$A^{~\,b}_{j\,\n}$}
\Text(90,20)[c]{$=$}
\Text(100,20)[l]{$ \ds
    {\d^{ab}\La^4 \over p^2\,(p^6+p^2\La^4+M_j^2\La^4)} \>
    \Big( \, p^2 g_{\m\n} -\; p_\m p_\n \,\Big) $}
\end{picture}\end{center}
\begin{center}\begin{picture}(330,40)(0,0)
\Line(10,21)(40,21)\Line(10,19)(40,19)\Line(40,19)(40,21)
\Gluon(40,20)(70,20){4}{3}
\Vertex(10,20){2}\Vertex(70,20){2}
\Text(10,12)[tc]{$b^a_j$} \Text(40,30)[bc]{${\ds p\atop \to}$}
\Text(70,12)[tc]{$A^{~\,a}_{j\,\m}$}
\Text(90,20)[c]{$=$}
\Text(100,20)[l]{$ \ds -\,i\,\d^{ab}~{p_\m \over p^2} $}
\end{picture}\end{center}
\begin{center}\begin{picture}(330,40)(0,0)
\Line(10,21)(70,21)\Line(10,19)(70,19)
\Vertex(10,20){2}\Vertex(70,20){2}
\Text(10,12)[tc]{$b^a_j$} \Text(40,30)[bc]{${\ds p\atop \to}$}
\Text(70,12)[tc]{$b^b_j$}
\Text(90,20)[c]{$=$}
\Text(100,20)[l]{$ \ds -\, \d^{ab}\> {M_j^2 \over p^2} $}
\end{picture}\end{center}
\begin{center}\begin{picture}(330,40)(0,0)
\DashArrowLine(10,20)(70,20){2}\Vertex(10,20){2}\Vertex(70,20){2}
\Text(10,12)[tc]{$\barc^a_i$} \Text(40,30)[bc]{$p$}
\Text(70,12)[tc]{$c^b_i$}
\Text(90,20)[c]{$=$}
\Text(100,20)[l]{$ \ds -\, {\d^{ab} \over p^2 + m_i^2} ~, $}
\end{picture}\end{center}
where $p^4$ stands for $(p^2){}^2.$ Note that the auxiliary field
$b^a_j$ propagates with propagator proportional to $M_j^2,$ whereas
the field $b^a$ does not propagate at all. As for the vertices, we
group them according to their order in the coupling constant $g.$
Vertices of order $g$ are given by
\begin{center}\begin{picture}(410,105)(0,0)
\Photon(0,22.5)(30,40){2}{4} \LongArrow(4.5,33)(14,39)
\Photon(30,40)(60,22.5){2}{4} \LongArrow(46.5,22)(37,28)
\Photon(30,40)(30,75){2}{4} \LongArrow(36,64)(36,53)
\Vertex(30,40){2}
\Text(30,86)[cb]{$A^{a_1}_{\m_1}$} \Text(39,59)[cl]{$p_1$}
\Text(3,14)[tr]{$A^{a_2}_{\m_2}$} \Text(8,40)[cr]{$p_2$}
\Text(55,14)[tl]{$A^{a_3}_{\m_3}$} \Text(45,20)[tr]{$p_3$}
\Text(85,47.5)[c]{$=$}
\Gluon(110,22.5)(140,40){3}{4}\Gluon(140,40)(170,22.5){3}{4}
\Photon(140,40)(140,75){2}{4}\Vertex(140,40){2}
\Text(140,86)[cb]{$A^{a_1}_{\m_1}$}
\Text(120,14)[tr]{$A_j{}^{a_2}_{\m_2}$}
\Text(160,14)[tl]{$A_j{}^{a_3}_{\m_3}$}
\Text(195,47.5)[c]{$=$}
\Text(210,62.5)[l]{$ {\ds {ig\over\La^4}\> {\cal S}_3 \Big\{
       f^{a_1a_2a_3}\, \Big[ \,\La^4 \, p_{1\m_2}\, g_{\m_3\m_1} \!
           - p_1^4 \, p_{1\m_2}\,g_{\m_3\m_1}  }$}
\Text(228,32.5)[l]{$ {\ds  +\, p_1^2\, (p_3\!-p_1)_{\m_2} \, \big(\,
       p_{1\m_3} \, p_{3\m_1}\! - p_1\!\!\cdot\! p_3\, g_{\m_3\m_1}
           \big) \,\Big] \Big\} }$}
\end{picture}\end{center}
\begin{center}\begin{picture}(187,85)(0,0)
\Line(0,23.8)(30.5,41.3)\Line(1.2,22)(30,38.3)\Line(1.2,22)(0,23.8)
\Gluon(30,40)(60,22.5){3}{4}
\Photon(30,40)(30,75){2}{4}\Vertex(30,40){2}
\Text(30,86)[cb]{$A^a_\m$}\Text(3,14)[tr]{$b^b_j$}
\Text(50,14)[tl]{$A_j{}^c_\n$}
\Text(85,47.5)[c]{$=$}
\Text(105,47.5)[l]{$rg f^{abc} g_{\m\n}\>,$}
\end{picture}\end{center}
\begin{center}\begin{picture}(187,85)(0,0)
\DashArrowLine(0,22.5)(30,40){6}\DashArrowLine(30,40)(60,22.5){6}
\Photon(30,40)(30,75){2}{4}\Vertex(30,40){2}
\Text(30,86)[cb]{$A^a_\m$} \Text(3,14)[tr]{$\barc^b$}
\Text(55,14)[tl]{$c^c$} \Text(48,34)[bl]{$p$}
\Text(85,47.5)[c]{$=$}
\Text(105,47.5)[l]{$-\,igf^{abc}\,p_\m$}
\end{picture}\end{center}
\begin{center}\begin{picture}(187,85)(0,0)
\DashArrowLine(0,22.5)(30,40){2}\DashArrowLine(30,40)(60,22.5){2}
\Photon(30,40)(30,75){2}{4}\Vertex(30,40){2}
\Text(30,86)[cb]{$A^a_\m$}\Text(3,14)[tr]{$\barc^b_i$}
\Text(55,14)[tl]{$c^c_i$}\Text(12,34)[br]{$p_2$}\Text(48,34)[bl]{$p_3$}
\Text(85,47.5)[c]{$=$}
\Text(105,47.5)[l]{$-\,igf^{abc}\,(\,rp_2+p_3 )_\m$}
\end{picture}\end{center}
\vspace{-10pt}
where ${\cal S}_3$ is the symmetrization operator with respect to the
indices 1, 2 and 3, and $r$ is the parameter distinguishing gauge
non-invariant $(r=0)$ from gauge invariant $(r=1)$ Pauli-Villars
determinants. Vertices of order $g^2$ have the form
\begin{center}\begin{picture}(170,90)(0,0)
\Photon(0,0)(25,25){2}{4}\LongArrow(10,3)(18,11)
                         \Text(20,2)[c]{$p_2$}
\Photon(25,25)(0,50){2}{4}\LongArrow(3,40)(11,32)
                         \Text(2,30)[c]{$p_1$}
\Photon(25,25)(50,0){2}{4}\LongArrow(47,10)(39,18)
                         \Text(52,18)[c]{$p_3$}
\Photon(25,25)(50,50){2}{4}\LongArrow(40,47)(32,39)
                         \Text(30,48)[c]{$p_4$}
\Vertex(25,25){2}
\Text(3,60)[br]{$A^{a_1}_{\m_1}$}\Text(3,-4)[tr]{$A^{a_2}_{\m_2}$}
\Text(45,-4)[tl]{$A^{a_3}_{\m_3}$}\Text(45,60)[bl]{$A^{a_4}_{\m_4}$}
\Text(80,25)[c]{$=$}
\Gluon(110,0)(135,25){3}{4}\Gluon(135,25)(160,0){3}{4}
\Photon(135,25)(110,50){2}{4}\Photon(135,25)(160,50){2}{4}
\Vertex(135,25){2}
\Text(117,60)[br]{$A^{a_1}_{\m_1}$}
\Text(120,-4)[tr]{$A_j{}^{a_2}_{\m_2}$}
\Text(150,-4)[tl]{$A_j{}^{a_3}_{\m_3}$}
\Text(155,60)[bl]{$A^{a_4}_{\m_4}$}
\Text(190,25)[c]{$=$}
\end{picture}\end{center}
\begin{displaymath}\begin{array}{l} {\ds
   \quad~ = -\,{g^2\over \La^4}\,{\cal S}_4 \Big\{
           f^{a_1a_2b}f^{a_3a_4b}\, \Big[ \,
              \La^4\,g_{\m_1\m_3} \, g_{\m_2\m_4}
            + (p_1\!+p_2)^4 \,g_{\m_1\m_3}\,g_{\m_2\m_4} } \\[9pt]
\phantom{\qquad\qquad\quad} {\ds
   +\, 4\,p_1^2\,g_{\m_2\m_3}\, \big( p_{4\m_1}\,p_{1\m_4} \!
           + p_1\!\!\cdot\! p_4\, g_{\m_1\m_4} \big)
   -\, 4\,p_1^2\,g_{\m_1\m_4}\,(2\,p_1 +p_2)_{\m_2} \,p_{1\m_3}
                                                         }\\[9pt]
\phantom{\qquad\qquad\quad} {\ds
    + \,2\,p_{1\m_1}\,p_{3\m_3}\,\big( p_2\!\cdot\! p_4\,g_{\m_2\m_4}
           - p_{2\m_4}\,p_{4\m_2} \big)
    +\, 4\,(p_1+p_2)^2\,g_{\m_2\m_4}\,p_{4\m_1}\,(p_3+2\,p_4)_{\m_3}
                                                         }\\[9pt]
\phantom{\qquad\qquad\quad} {\ds
    +\, 8\,(p_1\!+p_2)_{\m_1} \,p_{4\m_3} \big(
        p_2\!\cdot\! p_4\,g_{\m_2\m_4} \!- p_{2\m_4}\,p_{4\m_2} \big)
    \Big]\Big\} }
\end{array}\end{displaymath}
\begin{center}\begin{picture}(260,90)(0,0)
\DashArrowLine(0,0)(25,25){2}\Photon(25,25)(0,50){2}{4}
\DashArrowLine(25,25)(50,0){2}\Photon(25,25)(50,50){2}{4}
\Vertex(25,25){2}
\Text(3,60)[br]{$A^{a_1}_\m$}\Text(3,-4)[tr]{$\barc^{\>a_2}_i$}
\Text(45,-4)[tl]{$c^{a_3}_i$}\Text(45,60)[bl]{$A^{a_4}_\n$}
\Text(80,25)[c]{$=$}
\Text(105,25)[l]{$g^2\, ( f^{a_1a_2b} f^{a_4a_3b}
                       + f^{a_1a_3b} f^{a_4a_2b} ) \> g_{\m\n}$}
\end{picture}\end{center}
where ${\cal S}_4$ is the symmetrization operator with respect to the
indices 1, 2, 3 and 4. The action $S_\La$ and $\det{\BB A}_j$ also
contain vertices of higher order in $g.$ However, these only enter
two and higher-loop 1PI diagrams, so we do not need them here.
}

\vspace{24pt}

\noindent{\large\bf Appendix B:~~ Computing the limit
$\La,M_j,m_i\to\infty$}\vspace{8pt}
{\renewcommand{\theequation}{B.\arabic{equation}}\setcounter{equation}{0}

\noindent
In this Appendix we give the details of the calculations leading to
the results of sections 4 and 5. We start presenting a large mass
vanishing theorem which simplifies very much the computation of the
large mass limit of the Feynman integrals we are interested in. Let us
denote by $p^\m_1,\ldots,p^\m_E$ some external momenta lying in a
bounded subdomain of ${\BB R}^n$ and let $m>0$ be a mass. Consider an
$L\!$-loop integral of the form
\begin{displaymath}
  I(p,m) \,= m^{\eta} \int \! d^n\!q_1\ldots d^n\!q_L ~
  {M(q) \over \prod_i \big[\, (k_i^2)^{\,n_i}
                          + m_i^{2n_i} \big]^{l_i}} ~,
\end{displaymath}
where $\eta$ is an arbitrary real number and
\begin{equation}
\begin{array}{l}
  M(q) =\, {\rm monomial~in~the~components~of~}
                      q_1^\m,\ldots,q^\m_L  \\
  k_i^\m =\, {\rm linear~combination~of~} p_1^\m,\ldots,p_E^\m
                     ~{\rm and}~ q^\m_1,\ldots,q^\m_L   \\
  m_i =\, {\rm either}~0~{\rm or}~m \\
  n_i,l_i =\, {\rm positive~integers}\,.
\end{array}
\nonumber
\end{equation}
Given a subintegral $J$ of $I(p,m),$ we call $\underline{\om}_{\,J}$
to its infrared degree at vanishing external momenta and denote by
$\underline{\om}_{\,\rm min} = {\rm min}_{\,J}\,\big\{0,
v\underline{\om}_{\,J}\big\}$ the minimum of zero and the infrared
degrees $\underline{\om}_J$ of all the subintegrals $J$ of $I(p,m)$
including $I(p,m)$ itself. Then, the following theorem holds:

{\leftskip=1 true cm \rightskip=1 true cm
\noindent {\it $m\!$-Theorem.} If the integral $I(p,m)$ is absolutely
convergent at non-exceptional external momenta, and if its mass
dimension $d$ and $\underline{\om}_{\rm min}$ defined above satisfy
$\,d - \underline{\om}_{\,\rm min} < 0,$ then $I(p,m)\to 0$ as
$m\to\infty.$
\par}

\noindent This is a trivial generalization of the $m\!$-theorem in
ref. \cite{largem}, to where we refer for the proof. Armed with this
theorem, we proceed to derive the results in eqs. (\ref{PiLM}),
(\ref{PiML}) and (\ref{PiL}) for the vacuum polarization tensor.

We start by recalling that the vacuum polarization tensor receives
contributions at one loop from the eleven Feynman diagrams in Figs. 1
and 2. To compute the diagrams, we first replace each diagram by its
dimensionally regularized version, then make our computations using
dimensional regularization techniques, and finally take the limit
$D\to 4.$ To see that this way to proceed is legitimate and does not
introduce ambiguities of any type, we recall the following property of
dimensionally regularized integrals \cite{Collins}: If $I$ is an
absolutely convergent integral defined in four dimensions, and if
$I(D)$ denotes the corresponding dimensionally regularized integral,
then
\begin{displaymath}
   I = \lim_{D\to 4} I(D) \,.
\end{displaymath}
This property, together with the fact that the sums $\s_1,~\s_2$ and
$\s^1_4:={\rm (2a)}$ in section 2 are given by sums of finite by power
counting integrals, implies that $\s_1,~\s_2$ and $\s_4^1$ can be
computed as explained. In the case of $\s_3,$ proving finiteness
required an additional regulator $\cR.$ Our approach here takes this
regulator to be dimensional regularization. It will of course happen
that isolated diagrams will have poles at $D=4,$ but these poles will
cancel after summing over diagrams and imposing conditions
(\ref{PValpha}) and (\ref{PVgamma}).

We first compute those diagrams which depend at most on one mass. To
illustrate how to proceed, we take diagram (1a) as an example and look
at $\a=0$ contributions. With the labeling of internal momenta in Fig.
1a, and dropping from the notation the delta $\d^{ab},$ the most
general dimensionally regularized integral arising from diagram (1a)
with $\a=0$ will have the form
\begin{equation}
   I^{\oa}_{\m\n}(D;n_q,n_k,n_\La) = \iddqgcv { \left( \,\La^{n_\La}
                \,q^{n_q} \,k^{n_k}\,p^{n_p} \, \right)_{\m\n}
         \over q^4\,(q^4+\La^4)\>k^4\,(k^4+\La^4) } ~.
\label{1ageneral}
\end{equation}
The numerator in the integrand is a monomial in $\La$ and in the
components of $q^\m,~k^\m\!=q\m\!+\!p^\m$ and $p^\m$ of degrees
$n_\La,~n_q,~n_k$ and $n_p,$ respectively. Since the vacuum
polarization tensor has mass dimension two, the $n{\rm 's}$ satisfy
$n_\La+n_q+n_k+n_p=14.$ Furthermore, since the propagator of the gauge
field carries two powers of the momentum in the numerator, we have
$n_q,n_k \geq 2,$ thus ensuring IR convergence. If $n_q+n_k<12,$ the
integral $I^{\oa}_{\m\n}(D;n_q,n_k,n_\La)$ is absolutely convergent by
power counting at $D=4.$ In this case we can set $D=4$ in eq.
(\ref{1ageneral}) and move on to computing its large-$\!\La$ limit. To
do the latter, we use the $m\!$-theorem above. It is trivial to see
that the theorem sates in this case that
\begin{displaymath}
   \left. {\begin{array}{c} n_q+n_k+n_\La < 12 \\
                         n_k+n_\La < 8 \end{array}} \right\}
   \qquad \Rightarrow \qquad
   \lim_{\La\to\infty} I^{\oa}_{\m\n}(4\,;n_q,n_k,n_\La)=0 \,.
\end{displaymath}
If $n_q+n_k+n_\La \geq 12$ and/or $n_k+n_\La \geq 8,$ we proceed
as follows. Take \eg\ the integral
\begin{equation}
  J^{\oa}_{\m\n}(D) = \iddqgcv { q^{10}\,k^2\,p_\m q_\n
         \over q^4\,(q^4+\La^4)\>k^4\,(k^4+\La^4) } ~.
\label{example1a}
\end{equation}
Substituting $k^\m$ by $q\m\!+\!p^\m$ and performing some trivial
algebra, we write
\begin{displaymath}
  J^{1a}_{\m\n}(D) = J'_{\m\n}(D) + J''_{\m\n}(D) \>,
\end{displaymath}
where
\begin{displaymath}
  J'_{\m\n}(D) = \iddqgcv { q^2 ( q^2 \!+ 2\,pq)\,p_\m q_\n
       \over (q^4+\La^4) ~ [\,(q+p)^4 +\La^4\,] }
\end{displaymath}
and
\begin{displaymath}
  J''_{\m\n}(D) = -\,\iddqgcv { q^2\,[ p^2 q^2
       - 2\,pq\,( p^2 + 2\,pq)\,] ~p_\m q_\n
       \over (q^4+\La^4) ~ (q+p)^2 ~ [\,(q+p)^4 +\La^4\,] } ~ .
\end{displaymath}
The integral $J''_{\m\n}(D)$ is absolutely convergent by power
counting at $D=4,$ so we can set $D=4$ in it. Doing so and using
the $m\!$-theorem, we have
\begin{displaymath}
  \lim_{\La\to\infty} J''_{\m\n}(4) = 0\,.
\end{displaymath}
To calculate $J'_{\m\n}(D),$ we use eq. (\ref{idfour}) iteratively and
whenever we come across an integral which is absolutely convergent by
power counting at $D=4,$ we set $D=4$ in it and use the $m\!$-theorem.
After two or three iterations, we get
\begin{displaymath}
  J'_{\m\n}(D) = \iddqgcv \bigg[\,
     {q^4 p_\m q_\n \over (q^4+\La^4)^2} -
     {2 q^2(pq)\,p_\m q_\n \over (q^4+\La^4)^2} -
     {4 q^6 (pq)\,p_\m q_\n \over (q^4+\La^4)^3}\,\bigg] + \ldots ~,
\end{displaymath}
with the dots collecting integrals which are finite at $D=4$ and whose
large-$\!\La$ limit at $D=4$ vanishes. Using eq. (\ref{dimregfour})
below, it is now trivial to see that
\begin{displaymath}
  J'_{\m\n}(D) = {3\over 2}\>\gtwo \> \bigg[\,{1\over \ee}
     + \ln\!\bigg({\La^2\over 4\pi\m^2}\bigg)
     + \,{1\over 6}\,\bigg]\,p_\m p_\n
\end{displaymath}
as $\ee\to 0,$ where $D=4+2\ee.$ In what follows we will denote by $v_0$
all mass-independent contributions which are finite at $D=4.$ With
this convention and the notation in eq. (\ref{Lz}), we finally write
\begin{displaymath}
   J^{1a}_{\m\n}(D) = {3\over 2}\>\gtwo\> L(\La^2)\> p_\m p_\n + \,v_0
   \>.
\end{displaymath}
The contribution of any other integral (\ref{1ageneral}) with
$n_q+n_k+n_\La\geq 12$ and/or $n_k+n_\La\geq 8$ can be calculated in
the same way. After some lengthy calculations and putting together all
contributions\footnote{The $\a\!$-dependent part of the diagram is
built up of integrals finite by power counting at $D=4$ whose
large-$\!\La$ limit can be computed using similar methods.}, we obtain
for the complete diagram:
\begin{equation}
\begin{array}{l}  {\ds
    {\rm (1a)} = \, \gtwo \, \bigg\{ \,{3\pi\over 2}\,
                       \bigg({\a\over 4} -5 \bigg) \La^2 g_{\m\n}
    - \,\bigg[\,{11\over 4}\,L(\La^2)
        + {1\over 2}\,\bigg( {25\over 6}-\a\bigg)
          \ln\!\bigg(\poLa\bigg) \bigg]\,p^2 g_{\m\n}  }\\[12pt]
\phantom{ {\ds {\rm (1a)}=\,\gtwo\,\bigg\{ \> }} {\ds
    - \, \bigg[ 13\,L(\La^2) - \bigg({7\over 3} - {\a\over 2}\bigg)
          \ln\!\bigg(\poLa\bigg) \bigg] \, p_\m p_\n \bigg\}
    +\, v_0 \>.}
\end{array}
\label{1a}
\end{equation}
Diagrams (1c), (1e), (1f), (2b) and (2e) are computed in the same way.
For them we obtain:
\begin{eqnarray}
    && \hspace{-20pt}
      {\rm (1c)} =\, \gtwo \, \bigg\{\,
      {3\pi\over 2}\,\bigg(5-{\a\over 4}\bigg)\,\La^2 g_{\m\n} \,
      + {4\over 3}\,L(\La^2)\,\Big( 11\,p^2 g_{\m\n}
                        + p_\m p_\n \Big) \bigg\}\! + v_0
\label{1c} \\[12pt]
    && \hspace{-20pt}
       {\rm (1e)} = -\,\gtwo \>L(p^2)\, \bigg(
       {1\over 12}\,p^2 g_{\m\n} + {1\over 6}\,p_\m p_\n \bigg)
       + \, v_0
\label{1e} \\[12pt]
    && \hspace{-20pt}
       {\rm (1f)} = -\,\gtwo  \, \sumi \ga_i \,L(m_i^2)\, \left[\,
            {(1+r)^2\over 2} \,\m_i^2 g_{\m\n}
          + {(1+r)^2\over 12}\> p^2 g_{\m\n}
          + {1-4r +r^2\over 6} \, p_\m p_\n \right] +\, v_0
\nonumber\\
    && \label{1f}
         \\
    && \hspace{-20pt}
        {\rm (2b)} = \, \gtwo \> r^2 \sumj \a_j
       \,L(p^2)\, \bigg( {1\over 12} \,p^2 g_{\m\n}
       + {1\over 6}\,p_\m p_\n \bigg) +\,v_0
\label{2b}\\[12pt]
    && \hspace{-20pt}
        {\rm (2e)} = \,\gtwo\> 2r \sumi \ga_i \, L(m^2_i) \>
         \m_i^2 g_{\m\n} \,+\, v_0 \>.
\label{2e}
\end{eqnarray}

Let us next look at diagrams (1b), (1d), (2a), (2c) and (2d). They all
depend on the masses $\La$ and $M_j$ and their limit $\La,M_j\to
\infty$ is not unique but depends on the way in which we approach
$\La=M_j=\infty.$ The three cases considered in section 4 are:

\bigskip
\noindent{\underline{\sl Case $\th_j=\infty$}}\medskip

\noindent
This corresponds to taking $M_j\to\infty ~(\La<\infty$ fixed) first
and then sending $\La\to \infty.$ Let us consider diagram (1b). The
most general integral it gives rise to has the form
\begin{equation}
   I^{\ob}_{\m\n}(D;n_q,n_k,n_\La) = \sumj \,\a_j ~\iddqgcv
      { \left( \,\La^{n_\La} \,q^{n_q} \,k^{n_k}\,p^{n_p}
         \,\right)_{\m\n}\over q^2\,D_j(q) \,k^4\, D_j(k) } ~,
\label{1bgeneral}
\end{equation}
with
\begin{displaymath}
  D_j(q)= q^6 + q^2 \La^4 + M_j^2 \La^4 \,.
\end{displaymath}
Again we have that $n_\La+n_q+n_k+n_p=14$ and $n_q,n_k \geq 2.$ If
$n_q+n_k<12,$ the integral is absolutely convergent by power counting
at $D=4$ and we have that
\begin{equation}
  0 \leq \Bigl|\, I^{\ob}_{\m\n}(4;n_q,n_k,n_\La) \,\Bigr|
    \leq g^2 \cv  \sumj \big| \a_j \big| \idq { \left(\,\La^{n_\La}\,
           |q|^{n_q}\,|k|^{n_k}\,|p|^{n_p}\,\right)_{\m\n}
           \over q^2\,(q^6+\la_j^6)~ k^4\, (k^6+\la_j^6) } ~,
\label{bound}
\end{equation}
where
\begin{displaymath}
\la_j = M_j^{1/3}\La^{2/3}\,.
\end{displaymath}
The $m\!$-theorem above implies that the the right-hand side in eq.
(\ref{bound}) goes to zero as $M_j$ approaches infinity and $\La$ is
kept fixed, thus yielding
\begin{displaymath}
  n_q+n_k <12 \qquad \Rightarrow \qquad
  \lim_{M_j\to\infty} I^{\ob}_{\m\n}(4\,;n_q,n_k,n_\La) =0 \,.
\end{displaymath}
To illustrate how to proceed if $n_q+n_q\geq 12,$ let us consider the
integral [compare with eq. (\ref{example1a})]
\begin{equation}
  J^{\ob}_{\m\n}(D) = \sumj \,\a_j ~ \iddqgcv { q^{10}\,k^2\,p_\m q_\n
         \over q^2\,D_j(q) \>k^2\,D_j(k) } ~.
\label{example1b}
\end{equation}
Substituting $k^\m$ by $q^\m\!+p^\m$ and using the identity
\begin{equation}
{1 \over D_j(q+p)} = {1\over D_j(q)}
                   - { D_j(q+p) - D_j(q) \over D_j(q)\> D_j(q+p)}
\label{idsixqp}
\end{equation}
and the $m\!$-theorem, we get
\begin{displaymath}
  J^{\ob}_{\m\n}(D) = \sumj \, \a_j ~ \iddqgcv
     {q^8\, p_\m q_\n \over D^2_j(q)} \, \bigg[ \,1 -
     {6\,q^4\,(pq) \over D_j(q)}\, \bigg] + \ldots
\end{displaymath}
where the dots stand for integrals finite by power counting at $D=4$
that vanish at $D=4$ as $M_j\to\infty$ and $La<\infty$ is kept fixed.
Recalling now
\begin{displaymath}
  {1\over D_j(q)} =  {1\over q^6 + \la_j^6} -
      {q^2 \La^4\over (q^6 + \la_j^6)\> D_j(q)}
\end{displaymath}
and invoking the $m\!$-theorem again, we have
\begin{displaymath}
  J^{\ob}_{\m\n}(D) = \sumj \, \a_j ~ \iddqgcv
     {q^8 p_\m q_\n \over (q^6+\la_j^6)^2} \bigg[ \,1 -
     {6\,q^4\, (pq)\over (q^6+\la_j^6)}\, \bigg]
     + \ldots
\end{displaymath}
Peforming the integral with the help of eq. (\ref{dimregsix}), we
finally obtain
\begin{displaymath}
  J^{\ob}_{\m\n}(D) = {3\over 2} ~ \gtwo \, \sumj \a_j \, L(\la_j^2)
      \> p_\m p_\n + v_0 \,,
\end{displaymath}
where the notation for $L(\la_j^2)$ is as in eq. (\ref{Lz}). Any other
integral (\ref{1bgeneral}) with $n_q+n_k\geq 12$ can be evaluated in a
similar fashion. After doing so and summing over all contributions to
diagram (1b), we finally get
\begin{equation}
  {\rm (1b)} = -\,\gtwo \,\sumj \a_j \,\bigg[\,
     4\pi\sqrt{3}~\la_j^2 g_{\m\n} +
     L(\la_j^2) \> \bigg( {11\over 4}\,p^2 g_{\m\n}+ 13\,p_\m p_\n
                     \bigg)\,\bigg] +\, v_0 \,.
\label{1bLM}
\end{equation}
Note that, provided one imposes the Pauli-Villars condition
(\ref{PValpha}), the pole in $L(\la_j^2)$ cancels with the pole from
$L(\La^2)$ in eq. (\ref{1a}) so as to give a finite partial sum
$\s_1.$

The other diagrams are evaluated in the same way. For themOB we obtain:
\begin{eqnarray}
   && \hspace{-30pt}
      {\rm (1d)} = {1\over 3}\> \gtwo \sumj \a_j \,\bigg[\,
     {23\pi\sqrt{3}\over 2}\> \la_j^2 g_{\m\n} +
     4\,L(\la_j^2) \> \Big( 11\,p^2 g_{\m\n}+ p_\m p_\n \Big)\bigg]
     + v_0  \label{1dLM} \\[12pt]
   && \hspace{-30pt}
      {\rm (2a)} = {r^2\over 3} \> \gtwo \sumj \a_j \,\bigg[
     - {\pi\sqrt{3}\over 2} \> \la_j^2 g_{\m\n}
     + \ln\!\bigg({p^2\over \la_j^2}\bigg)
        \bigg( {1\over 4} \,p^2 g_{\m\n} - p_\m p_\n \bigg)\bigg]
     + v_0  \label{2aLM} \\[12pt]
   && \hspace{-30pt}
      {\rm (2c)+(2d)} = {r\over 3} \> \gtwo \sumj \a_j \,\bigg[\,
     \pi\sqrt{3}\>\la_j^2 g_{\m\n} + 2\,L(\la_j^2) \,
      \bigg( {1\over 4}\,p^2 g_{\m\n} - p_\m p_\n \bigg)\bigg] +
     v_0\,. \label{2c2dLM}
\end{eqnarray}
Taking $r=1$ and summing eqs. (\ref{1a})-(\ref{2e}) and
(\ref{1bLM})-(\ref{2c2dLM}), we obtain eq. (\ref{PiLM}). Note that, in
agreement with our discussion in section 2, the partial sums
$\s_1,~\s_2,~\s_3^1$ and $\s_4^1:={\rm (2a)}$ are finite (no poles)
provided condition (\ref{PValpha}) and (\ref{PVgamma}) are met. Note
also that contributions of type $(masses)^2g_{\m\n}$ cancel in the
vacuum polarization tensor upon summation. If we take $r=0$ instead,
we recover the result in eq. (\ref{r0PiLM}).

\bigskip
\noindent{\underline{\sl Case $\th_j=0$}}\medskip

\noindent
This corresponds to first taking $\La\to\infty$ at fixed $M_j$ and
then sending $M_j$ to $\infty.$ Let us consider diagram (1b). As we
already know, the most general dimensionally regularized Feynman
integral arising from it has the form in eq. (\ref{1bgeneral}). If
$n_q+n_k<12,$ the integral is finite by power counting at $D=4.$ In
this case, the inequality
\begin{displaymath}
  0 \leq \Bigl|\, I^{\ob}_{\m\n}(4\,;n_q,n_k,n_\La) \,\Bigr| \leq \,
         g^2 \cv  \sumj \big| \a_j \big| \idq { \left(\,\La^{n_\La}\,
           |q|^{n_q}\,|k|^{n_k}\,|p|^{n_p}\,\right)_{\m\n}
           \over q^4\,(q^4+\La^4)~ k^4\, (k^4+\La^4) }
\end{displaymath}
and the $m\!$-theorem yield
\begin{equation}
   \left. {\begin{array}{c} n_q+n_k+n_\La < 12 \\
                         n_k+n_\La < 8 \end{array}} \right\}
   \qquad \Rightarrow \qquad
   \lim_{\La\to\infty} I^{\ob}_{\m\n}(4\,;n_q,n_k,n_\La)=0 \,.
\label{cut1bML}
\end{equation}
To explain how to proceed if $n_q+n_k+n_\La\geq 12$ and/or
$n_k+n_\La\geq 8,$ we take again the integral $J^{\ob}_{\m\n}(D)$ in
eq. (\ref{example1b}). Using eq. (\ref{idsix}), eq. (\ref{idsixqp})
and the $m\!$-theorem, it is very easy to see that
\begin{displaymath}
   I^{\ob}_{\m\n}(D) = \sumj\,\a_j \iddqgcv \bigg[\,
     {q^4 p_\m q_\n \over (q^4+\La^4)^2} -
     {3\,q^2 (pq)\>(2q^4+\La^4)\,p_\m q_\n \over (q^4+\La^4)^3}
     \,\bigg] + \ldots ~ ,
\end{displaymath}
with the dots collecting integrals finite by power counting at $D=4$
that at $D=4$ vanish as $\La\to\infty$ and $M_j$ is kept fixed.  To
perform the integration over $q^\m,$ we employ eq. (\ref{dimregfour})
below and obtain
\begin{displaymath}
   I^{\ob}_{\m\n}(D) = {3\over 2}\> \gtwo \> \sumj \, \a_j \,
      L(\La^2)\> p_\m p_\n + v_0.
\end{displaymath}
Proceeding in this way for all integrals (\ref{1bgeneral}) from
diagram (1b) that escape the cut (\ref{cut1bML}) and putting together
all contributions, we end up with
\begin{equation}\begin{array}{l} {\ds
  {\rm (\ob)} = \gtwo \,\sumj \a_j \,\bigg\{
      \bigg[{3\over 4}\, \ln\!\bigg(\poLa\bigg)
         - {9\over 4} \ln\!\bigg({\La^2\over M_j^2}\bigg)
         - {15 \over 2}\, \bigg] M_j^2 g_{\m\n}
}\\[12pt] \phantom{{\ds {\rm(1b) ~} }} {\ds
  - \,\bigg[ \, {11\over 4}\,L(\La^2)
         - {25\over 12}\,\ln\!\bigg({\La^2\over M_j^2}\bigg)\bigg]
                  p^2 g_{\m\n}
  - \bigg[ 13\,L(\La^2)
         + {7\over 3}\, \ln\!\bigg({\La^2\over M_j^2}\bigg)\bigg]
    p_\m p_\n \bigg\} + v_0 \,.}
\end{array}
\label{1bML}
\end{equation}
For the other diagrams we get
\begin{equation} \leftline{\quad~ ${\ds
   {\rm (1d)} = \gtwo \,\sumj \a_j \,\bigg\{ \bigg[\,
     {9\over 4}\,\ln\!\bigg({\La^2\over M_j^2}\bigg) +
             {15\over 2}\,\bigg] M_j^2g_{\m\n} +
     {4\over 3}\>L(\La^2)\,\Big( 11\,p^2g_{\m\n}
             + p_\m p_\n\Big) \bigg\} \!+ v_0 }$
\label{1dML}}
\end{equation}
\begin{equation} \leftline{\quad~ ${\ds
   {\rm (2a)} = r^2 \> \gtwo \,\sumj \a_j \,\bigg[\,
     {3\over 4}\, \ln\!\bigg(\poLa\bigg)\, M_j^2 g_{\m\n} +
     {1\over 3}\,\ln\!\bigg(\poMj\bigg)
        \bigg( {1\over 4} \,p^2 g_{\m\n} - p_\m p_\n \bigg)\bigg]\!
     + v_0 }$
\label{2aML}}
\end{equation}
\begin{equation}
\begin{array}{l} {\ds \hspace{-1pt}
     {\rm (2c)+(2d)} = \,r^2\>\gtwo \,\sumj \a_j \,\bigg\{ \!
     -{3\over 2}\>\ln\!\bigg(\poLa\bigg)\, M_j^2 g_{\m\n} }
   \\[12pt]
\phantom{ {\ds \hspace{-1pt} {\rm (2c)+(2d)} } } {\ds
   + \,{1\over 6}\, \bigg[ L(\La^2) +
      8\,\ln\!\bigg({\La^2\over M_j^2}\bigg)\bigg]\,
      p^2 g_{\m\n}
   -  {2\over 3}\, \bigg[ L(\La^2) +
      {5\over 4}\, \ln\!\bigg({\La^2\over M_j^2}\bigg)\bigg]\,
      p_\m p_\n \bigg\} \!+ v_0\,. }
\end{array}
\label{2c2dML}
\end{equation}
Eq. (\ref{PiLM}) follows from eqs. (\ref{1a})-(\ref{2e}) and
(\ref{1bML})-(\ref{2c2dML}). Note again that the sums $\s_l$ are
finite and that the resulting vacuum polarization tensor is
transverse.

\bigskip
\noindent{\underline{\sl Case $0<\th_j<\infty$}}\medskip

\noindent
Using the same type of arguments methods as for the previous two
cases, we have
\begin{eqnarray}
   && \hspace{-40pt}
      {\rm (1b)} = \gtwo \,\sumj \a_j \,\bigg[\, a_1\, \La^2 g_{\m\n}
    - L(\La^2)\,  \bigg( {11\over 4} \,p^2 g_{\m\n}
             + 13\,p_\m p_\n\bigg) \bigg] + \,v_0
\label{1bL}\\[12pt]
   && \hspace{-40pt}
      {\rm (1d)} = \gtwo \,\sumj \a_j \,\bigg[\, a_2 \,\La^2 g_{\m\n}
    + {4\over 3} \,L(\La^2)\,  \bigg( 11\,p^2 g_{\m\n}
             + p_\m p_\n\bigg) \bigg] + \,v_0
\label{1dL}\\[12pt]
   && \hspace{-40pt}
      {\rm (2a)} = r^2 \> \gtwo \,\sumj \a_j \,\bigg[\,
     a_3\, \La^2 g_{\m\n} + {1\over 3}\, \ln\!\bigg(\poLa\bigg)
        \bigg( {1\over 4} \,p^2 g_{\m\n} - p_\m p_\n \bigg)\bigg]
     + v_0
\label{2aL}\\[12pt]
   && \hspace{-40pt}
      {\rm (2c)+(2d)} = r^2\>\gtwo \,\sumj \a_j \,\bigg[ \,
     a_4\,\La^2 g_{\m\n}
     + {1\over 3}\> L(\La^2) \bigg( {1\over 2} \, p^2 g_{\m\n}
            - 2 p_\m p_\n \bigg) \bigg] + v_0 \,.
\label{2c2dL}
\end{eqnarray}
Here $a_1,\ldots,a_4$ are coefficients that depend on $\th_j$ and that
can be expressed in terms of the roots in the complex upper half-plane
of the polynomial $P(z)=z^6+z^2+\th_j^2.$

\bigskip
\noindent{\underline{\sl Useful integrals}}\medskip

\noindent
We finish this appendix by giving the following two dimensionally
regularized integrals that have been widely used in the computations
and that are not in the tables:
\begin{equation}
  \iddq { (q^2)^\eta \over (q^4 + \la^4)^\ga} \,=~
      {\la^{4+2\eta-4\ga}\over 32\pi^2}~
      \bigg({\la^2\over 4\pi\m^2}\bigg)^{\!{D\over 2} -2}\;
      {\gm\!\left({D\over 4} +{\eta\over 2}\right) \,
      \gm\!\left(\ga-{D\over4}-{\eta\over 2}\right)
                \over \gm(\ga)\> \gm\!\left({D\over2}\right) }
\label{dimregfour}
\end{equation}
\begin{equation}
   \iddq { (q^2)^\eta \over (q^6 + \la^6)^\ga} \,=~
     {\la^{4+2\eta-6\ga}\over 48\pi^2} ~
     \bigg({\la^2\over 4\pi\m^2}\bigg)^{\!{D\over 2}-2}\;
     {\gm\!\left({D\over 6}+{\eta\over 3}\right) \,
      \gm\!\left(\ga-{D\over 6}-{\eta\over 3}\right)
          \over \gm(\ga)\> \gm\!\left({D\over 2}\right)} ~.
\label{dimregsix}
\end{equation}

\bigskip\bigskip

%
\def\section{\subsection}

\newpage

\begin{center}
\begin{picture}(460,140)(0,0)
\Photon(20,100)(60,100){3}{3}
\Photon(140,100)(180,100){3}{3}
\PhotonArc(100,100)(40,0,180){3}{8.5}
\PhotonArc(100,100)(40,180,360){3}{8.5}
\Vertex(60,100){2} \Vertex(140,100){2}
\Text(25,86)[l]{$A$} \Text(175,86)[r]{$A$}
\Text(62,64)[r]{$A$} \Text(138,64)[l]{$A$}
\Text(62,136)[r]{$A$} \Text(138,136)[l]{$A$}
\Text(100,20)[]{(a)}
\Photon(280,100)(320,100){3}{3}
\Photon(400,100)(440,100){3}{3}
\GlueArc(360,100)(40,0,180){4}{8}
\GlueArc(360,100)(40,180,360){4}{8}
\Vertex(320,100){2} \Vertex(400,100){2}
\Text(240,100)[l]{${\displaystyle \sum_j} ~\a_j$}
\Text(285,86)[l]{$A$} \Text(435,86)[r]{$A$}
\Text(322,64)[r]{$A_j$} \Text(400,64)[l]{$A_j$}
\Text(322,136)[r]{$A_j$} \Text(400,136)[l]{$A_j$}
\Text(360,20)[]{(b)}
\end{picture}
\end{center}

\vspace{0.6cm}

\begin{center}
\begin{picture}(460,120)(0,0)
\Photon(35,40)(165,40){3}{8.5} \PhotonArc(100,85)(40,270,269){3}{15.5}
\Vertex(100,43.5){2}
\Text(40,26)[l]{$A$} \Text(160,26)[r]{$A$}
\Text(52,85)[r]{$A$} \Text(148,85)[l]{$A$}
\Text(100,0)[]{(c)}
\Photon(295,40)(425,40){3}{8.5} \GlueArc(360,85)(40,-90,270){4}{16}
\Vertex(360,43.5){2}
\Text(255,40)[l]{${\displaystyle \sum_j} ~\a_j$}
\Text(300,26)[l]{$A$} \Text(420,26)[r]{$A$}
\Text(312,85)[r]{$A_j$} \Text(410,85)[l]{$A_j$}
\Text(360,0)[]{(d)}
\end{picture}
\end{center}

\vspace{0.6cm}

\begin{center}
\begin{picture}(460,160)(0,0)
\Photon(20,100)(60,100){3}{3}
\Photon(140,100)(180,100){3}{3}
\DashArrowArc(100,100)(40,0,180){10}
\DashArrowArc(100,100)(40,180,360){10}
\Vertex(60,100){2} \Vertex(140,100){2}
\Text(25,86)[l]{$A$} \Text(175,86)[r]{$A$}
\Text(68,64)[r]{$c$} \Text(136,64)[l]{$\bar{c}$}
\Text(68,136)[r]{$\bar{c}$} \Text(136,136)[l]{$c$}
\Text(100,20)[]{(e)}
\Photon(280,100)(320,100){3}{3}
\Photon(400,100)(440,100){3}{3}
\DashArrowArc(360,100)(40,0,180){2}
\DashArrowArc(360,100)(40,180,360){2}
\Vertex(320,100){2} \Vertex(400,100){2}
\Text(240,100)[l]{${\displaystyle \sum_j} ~\ga_i$}
\Text(285,86)[l]{$A$} \Text(435,86)[r]{$A$}
\Text(327,64)[r]{$c_i$} \Text(396,64)[l]{$\bar{c}_i$}
\Text(327,136)[r]{$\bar{c}_i$} \Text(396,136)[l]{$c_i$}
\Text(360,20)[]{(f)}
\end{picture}
\end{center}

\vspace{0.3cm}

\begin{tabbing}
\qquad \= {\bf Figure 1:} \= {\sl One-loop regularization of the
vacuum polarization tensor: diagrams} \kill
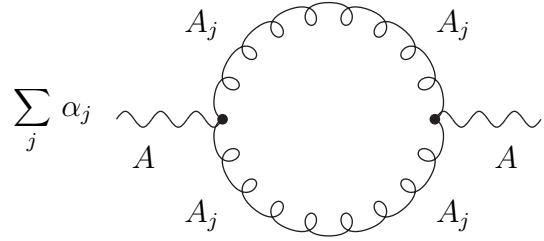
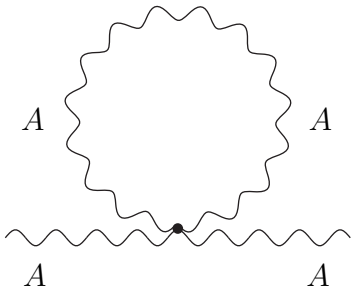
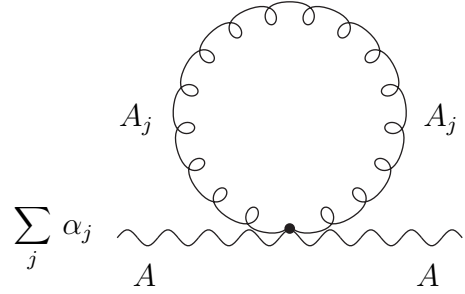
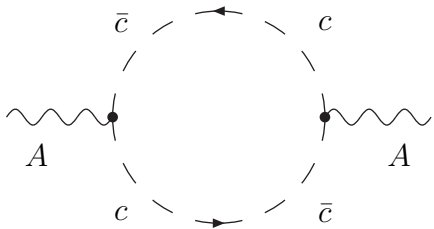
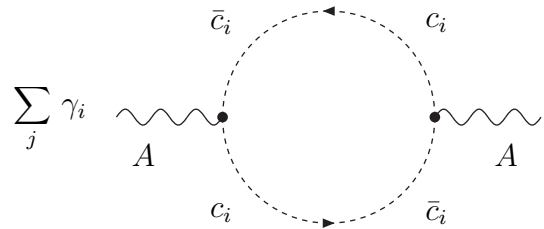
\> {\bf Figure 1:} \> {\sl One-loop regularization of the vacuum
polarization tensor: diagrams} \\
\>                 \> {\sl that contribute for both $r=0,1$. \,Recall
that the contribution of \,{\rm (f)}} \\
\>                 \> {\sl depends on the value of $r$.} \\
\end{tabbing}

\newpage

\begin{center}
\begin{picture}(460,140)(0,0)
\Photon(40,100)(80,100){3}{3}
\Photon(160,100)(200,100){3}{3}
\GlueArc(120,100)(40,0,180){4}{8}
\CArc(120,100)(38.5,180,360)\CArc(120,100)(41.5,180,360)
\Vertex(80,100){2} \Vertex(160,100){2}
\Text(0,100)[l]{${\displaystyle \sum_j} ~\a_j$}
\Text(45,86)[l]{$A$} \Text(195,86)[r]{$A$}
\Text(82,64)[r]{$b_j$} \Text(158,64)[l]{$b_j$}
\Text(82,136)[r]{$A_j$} \Text(158,136)[l]{$A_j$}
\Text(120,20)[]{(a)}
\Photon(300,100)(340,100){3}{3}
\Photon(420,100)(460,100){3}{3}
\CArc(380,100)(41.5,0,90)\CArc(380,100)(38.5,0,90)\Line(380,139)(380,141)
\GlueArc(380,100)(40,90,180){4}{4}
\CArc(380,100)(38.5,180,270)\CArc(380,100)(41.5,180,270)
\Line(380,59)(380,61)
\GlueArc(380,100)(40,270,360){4}{4}
\Vertex(340,100){2} \Vertex(420,100){2}
\Text(260,100)[l]{${\displaystyle \sum_j} ~\a_j$}
\Text(295,86)[l]{$A$} \Text(455,86)[r]{$A$}
\Text(342,64)[r]{$b_j$} \Text(420,64)[l]{$A_j$}
\Text(342,136)[r]{$A_j$} \Text(420,136)[l]{$b_j$}
\Text(380,20)[]{(b)}
\end{picture}
\end{center}

\vspace{0.9cm}

\begin{center}
\begin{picture}(460,140)(0,0)
\Photon(40,100)(80,100){3}{3}
\Photon(160,100)(200,100){3}{3}
\GlueArc(120,100)(40,0,180){4}{8}
\GlueArc(120,100)(40,180,270){4}{4}
\CArc(120,100)(41.5,270,360)\CArc(120,100)(38.5,270,360)
\Line(120,58.8)(120,61.5)
\Vertex(80,100){2} \Vertex(160,100){2}
\Text(0,100)[l]{${\displaystyle \sum_j} ~\a_j$}
\Text(45,86)[l]{$A$} \Text(195,86)[r]{$A$}
\Text(82,64)[r]{$A_j$} \Text(158,64)[l]{$b_j$}
\Text(82,136)[r]{$A_j$} \Text(158,136)[l]{$A_j$}
\Text(120,0)[]{(c)}
\LongArrowArc(120,100)(28,70,112) \Text(120,120)[tc]{$q$}
\LongArrowArc(120,73)(28,250,292) \Text(120,40)[tc]{$q+p$}

\Photon(300,100)(340,100){3}{3}
\Photon(420,100)(460,100){3}{3}
\CArc(380,100)(41.5,0,90)\CArc(380,100)(38.5,0,90)
\Line(380,139)(380,141)
\GlueArc(380,100)(40,90,180){4}{4}
\GlueArc(380,100)(40,180,360){4}{8}
\Vertex(340,100){2} \Vertex(420,100){2}
\Text(260,100)[l]{${\displaystyle \sum_j} ~\a_j$}
\Text(305,86)[l]{$A$} \Text(455,86)[r]{$A$}
\Text(342,64)[r]{$A_j$} \Text(420,64)[l]{$A_j$}
\Text(342,136)[r]{$A_j$} \Text(420,136)[l]{$b_j$}
\Text(380,0)[]{(d)}
\end{picture}
\end{center}

\begin{center}
\begin{picture}(460,130)(0,0)
\Photon(165,40)(295,40){3}{8.5} \DashArrowArc(230,82.5)(40,-90,270){2}
\Vertex(230,42.5){2}
\Text(125,40)[l]{${\displaystyle \sum_i}~\ga_i$}
\Text(170,26)[l]{$A$} \Text(290,26)[r]{$A$}
\Text(182,85)[r]{$\bar{c}_i$} \Text(278,85)[l]{$c_i$}
\Text(230,0)[]{(e)}
\end{picture}
\end{center}

\vspace{0.6cm}

\begin{tabbing}
\qquad \= {\bf Figure 2:} \= {\sl One-loop regularization of the
vacuum polarization tensor: diagrams} \kill
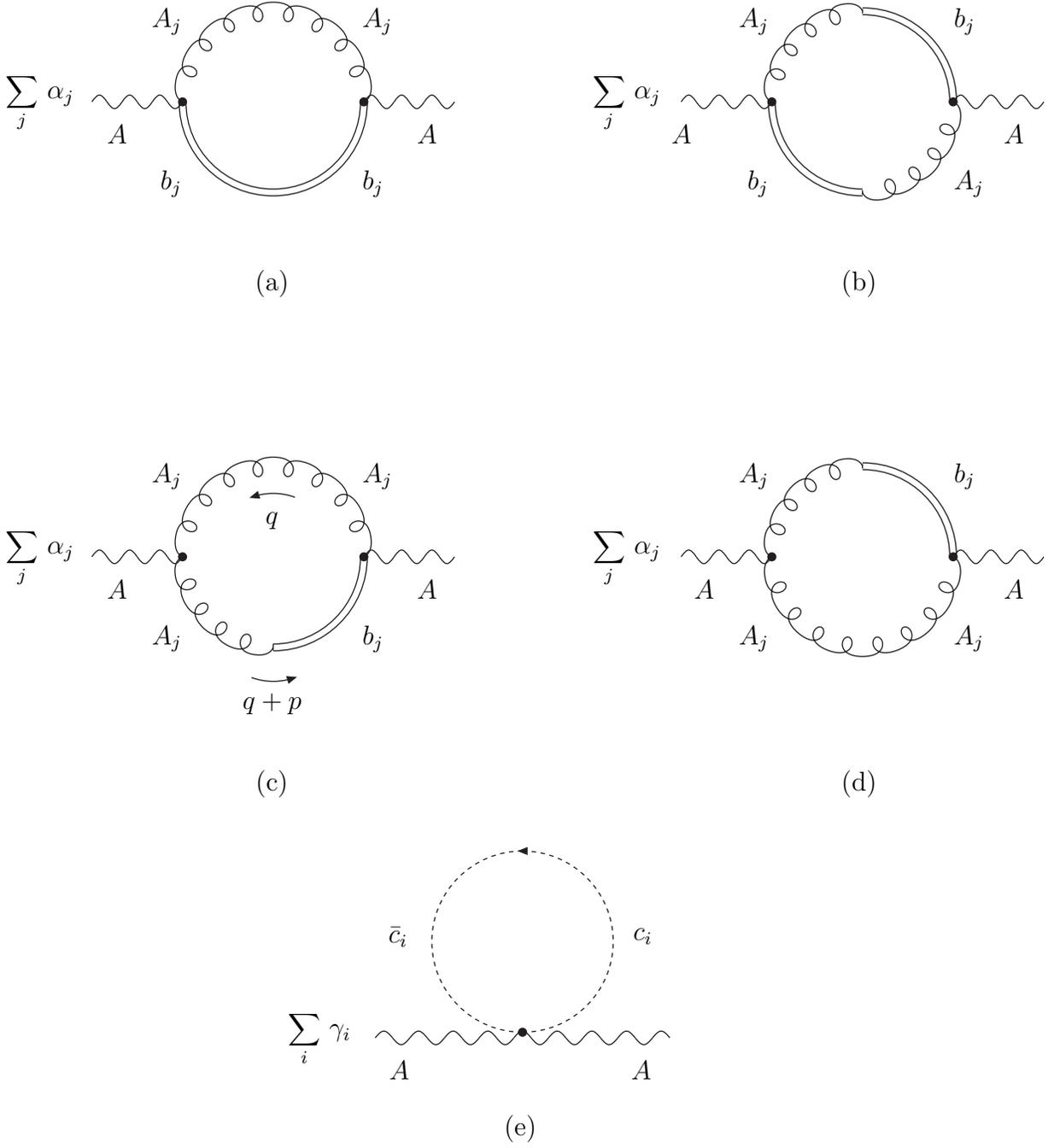
\> {\bf Figure 2:} \> {\sl One-loop regularization of the vacuum
polarization tensor: diagrams} \\
                \> \> {\sl that contribute only for $r=1$.} \\
\end{tabbing}

\vspace{0.9cm}
\begin{center}
\begin{picture}(200,60)(0,0)
\DashArrowLine(20,20)(60,20){6} \DashArrowLine(140,20)(180,20){6}
\DashArrowLine(60,20)(140,20){6}\PhotonArc(100,20)(40,0,180){3}{8.5}
\Vertex(60,20){2}\Vertex(140,20){2}
\Text(42,5)[b]{$c$} \Text(158,5)[b]{$\barc$}
\Text(78,5)[b]{$\barc$} \Text(122,5)[b]{$c$}
\Text(61,51)[cr]{$A$} \Text(140,51)[cl]{$A$}
\end{picture}
\end{center}
\begin{center}
\vspace{10pt}
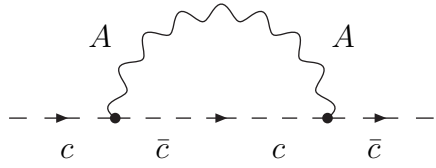
{\bf Figure 3:} {\sl One-loop regularization of the ghost
self-energy.}
\end{center}

\begin{center}
\begin{picture}(400,280)(0,0)
\DashArrowLine(15,15)(49.6,35){6}
\DashArrowLine(49.6,35)(84.2,95){6}
\DashArrowLine(84.2,95)(118.8,35){6}
\DashArrowLine(118.8,35)(153.4,15){6}
\Photon(49.6,35)(118.8,35){3}{5}\Photon(84.2,95)(84.2,135){3}{3}
\Vertex(49.6,35){2}\Vertex(118.8,35){2}\Vertex(84.2,95){2}
\Text(53.6,27)[tl]{$A$} \Text(112.8,27)[tr]{$A$}
\Text(32.6,32)[r]{$c$} \Text(139.8,32)[l]{$\barc$}
\Text(54.6,50)[r]{$\barc$} \Text(114.8,50)[l]{$c$}
\Text(73.2,83)[r]{$c$} \Text(97.2,83)[cl]{$\barc$}
\Text(89.2,115)[l]{$A$}
\DashArrowLine(250,15)(284.6,35){6}
\Photon(284.6,35)(319.2,95){3}{5}
\Photon(319.2,95)(353.8,35){3}{5}
\DashArrowLine(353.8,35)(388.4,15){6}
\DashArrowLine(284.6,35)(353.8,35){6}
\Photon(319.2,95)(319.2,135){3}{3}
\Vertex(284.6,35){2}\Vertex(353.8,35){2}\Vertex(319.2,95){2}
\Text(290.6,30)[tl]{$\barc$} \Text(347.8,27)[tr]{$c$}
\Text(267.6,32)[r]{$c$} \Text(374.8,32)[l]{$\barc$}
\Text(288.6,50)[br]{$A$} \Text(351.8,50)[bl]{$A$}
\Text(305.2,83)[r]{$A$} \Text(335.2,83)[cl]{$A$}
\Text(324.2,115)[l]{$A$}
\end{picture}
\end{center}
\begin{center}
\vspace{10pt}
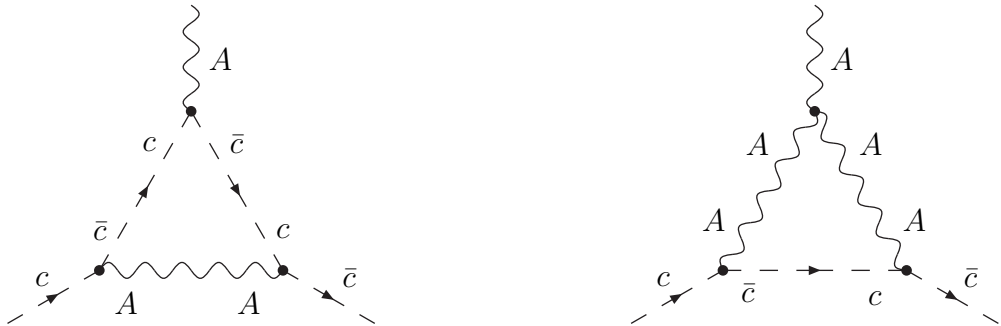
{\bf Figure 4:} {\sl One-loop regularization of the
ghost-ghost-vertex.}
\end{center}

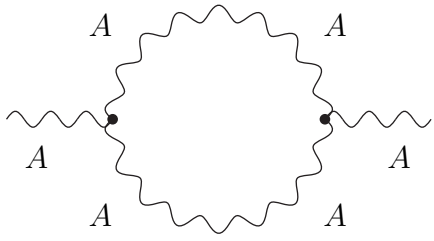
\end{document}